\documentclass[journal,romanappendices]{IEEEtran}
\usepackage{eurosym}
\usepackage{amsmath,amsthm}
\usepackage{amsfonts}
\usepackage{amssymb}
\usepackage{graphicx}
\usepackage{xcolor}
\usepackage{setspace}
\usepackage{amsmath}
\providecommand{\U}[1]{\protect\rule{.1in}{.1in}}
\providecommand{\U}[1]{\protect\rule{.1in}{.1in}}
\newtheorem{theorem}{Theorem}
\newtheorem{corollary}{Corollary}[theorem]
\providecommand{\U}[1]{\protect\rule{.1in}{.1in}}
\usepackage{cite}

\begin{document}

\begin{center}

{\large \textbf{Homogenizing Entropy Across \\Different Environmental
Conditions: \\A Universally Applicable Method \\for Transforming Continuous
Variables}}

\bigskip\bigskip\bigskip

{\large Joel R. Peck$^{1}$ and David Waxman$^{2}$} \bigskip

{\large $^{1}$ Department of Genetics,
University of Cambridge,
Cambridge CB2 3EH, UK\\
Email: jp564@cam.ac.uk }

\bigskip

{\large $^{2}$ Centre for Computational Systems Biology, \\ISTBI, Fudan
University, Shanghai 200433, PRC\\
Email: davidwaxman@fudan.edu.cn}
\end{center}

\newpage

\begin{center}
{\large \textbf{Abstract}}
\end{center}

In classical information theory, a causal relationship between two variables
is typically modelled by assuming that, for every possible state of one of the
variables, there exists a particular distribution of states of the second
variable. Let us call these two variables the \textit{causal} and
\textit{caused} variables, respectively. We shall assume that both variables
are continuous and one-dimensional. In this work we consider a procedure to
transform each variable, using transformations that are differentiable and
strictly increasing. We call these \textit{increasing transformations}. Any
causal relationship (as defined here) is associated with a channel capacity,
which is the maximum rate that information could be sent if the causal
relationship was used as a signalling system. Channel capacity is unaffected
when the two variables are changed by use of increasing transformations. For
any causal relationship we show that there is always a way to transform the
caused variable such that the entropy associated with the caused variable is
independent of the value of the causal variable. Furthermore, the resulting
universal entropy has an absolute value that is equal to the channel capacity
associated with the causal relationship. This observation may be useful in
statistical applications. For any causal relationship, it implies that there
is a `natural' way to transform a continuous caused variable. We also show
that, with additional constraints on the causal relationship, a natural
increasing transformation of both variables leads to a transformed causal
relationship that has properties that might be expected from a well-engineered
measuring device.\newline


\newpage

\section{Introduction}

\indent Many phenomena interact with one another. For example, objects
generally interact with light, and thus they affect the number and qualities
of the photons that they emit, absorb, or reflect. Because of these
interactions, various phenomena provide information about other phenomena.
Thus, for example, the pattern of light arriving from the direction of a
particular object will typically provide information about the properties of
that object. Nevertheless, if two phenomena interact and we try to use one to
assess the properties of the other, we are likely to find that the details of
the interaction make it difficult to produce a useful assessment. This is
because the interaction is unlikely to have the convenient characteristics
that are typical in the case of artificially created measuring instruments,
like thermometers, seismographs, Geiger counters, etc. Here, we show that it
is often possible to transform measurements of natural phenomena in a way that
gives them some (or all) of the sorts of convenient characteristics usually
associated with artificial measuring instruments. This may provide technical
advantages, and it also suggests, for a given interaction between two
phenomena, that there will often be natural transformed variables with which
it is convenient to measure these phenomena. This observation may prove useful
in a variety of contexts, including the measurement of biological adaptation
\cite{PeckWaxman2018}.

To begin, it is worth considering the case of an artificially designed
measuring apparatus. Let us say that we wish to measure the temperature of the
air in a particular room. Temperature (on, say, a Celsius scale) is a linear
function of the average kinetic energy of the particles that make up the air
in the room. As such, it has a definite value at any given time. Next, imagine
a digital thermometer that is situated in the room. Assume that this
thermometer has an extremely fine scale, so it divides each degree of
temperature into a very large number of equally sized parts.

The air in contact with the thermometer, at any given time, is only a small
sample of all the air in the room. For this reason (along with others) we
expect some difference between the reading of the thermometer and the actual
temperature of the air in the room. However, if the thermometer is functioning
properly then: (i) for a given temperature of the air in the room, the
readings of the thermometer should, typically, be approximately equal to the
actual temperature, (ii) the \textit{accuracy} and \textit{precision} of the
thermometer should roughly be the same for all temperatures within its
operating range. Here, accuracy refers to the proximity of
measured temperature to the actual temperature, while precision refers to the
extent to which repeated measurements, under the same conditions, yield
similar results.

The convenient characteristics of a typical thermometer are, of course, the
result of engineering efforts. Natural phenomena are usually very different.
For example, imagine a patch of ground where we notice that, after a
rainstorm, the patch tends to dry out faster on hot days than cold days. This
relationship may be fairly reliable, but it is unlikely to have the
characteristics of a good thermometer. For example, suppose we find that the
average drying time is approximately $200$ minutes at $10^{\circ}C$, and
approximately $100$ minutes at $20^{\circ}C$. \textit{If} drying time was like
a thermometer, then it would (approximately) be a linear function of
temperature, such that at $15^{\circ}C$ the drying time would be about $150$
minutes. Furthermore, the variation that occurs in drying time would be
approximately the same, regardless of whether the weather is warm or cold.
However, in reality, there is no reason to expect that drying time will
decrease linearly with temperature. For example, the effect of temperature on
drying time may be relatively small when the soil is already dry, and thus the
drying time, at $15^{\circ}C$, may substantially differ from $150$ minutes.
Similarly, variation in cloud cover may be much greater on cold days, and thus
variation in drying time may be much greater when the temperature is below
$10^{\circ}C$, as compared with warmer days. Can anything be done with natural
relationships so they can be made similar to the artificial relationships of
measuring devices that we design and manufacture?

As we shall see, it appears that the mathematical theory of information can
help. An example is the case of the relation between drying time and
temperature that is described above. In general, this relationship can be
expected to have some inconvenient features, including non-linearity. It may,
however, be possible to transform both of the variables (drying time and
temperature) to new variables, such that the new variable (that is a
transformation of the drying time) has a mean value that changes
\textit{strictly linearly} with the value of the other new variable (the
transformation of the temperature). Furthermore, these transformations may be
able to ensure that the level of variation in the transformed drying time will
always be the same, regardless of the value of the transformed temperature.

\section{Mathematical model}

\indent Let us consider a mathematical model in which the state of
a local environment is represented by a random variable $E$. We will use a
second random variable, $Q$, to represent the state (or quality) of some
system that will be used to measure the environment. For the sake of
simplicity, we will assume that both $E$ and $Q$ are continuously distributed
one-dimensional (or scalar) quantities, like length, mass, luminosity, etc. We
identify $E$ and $Q$ with the causal and caused variables, respectively, as
described above. 

Here, we present the principal results from the mathematical model. Some mathematical 
background of the 
distributions and related quantities, that appear in the model, are given in Appendix 
 \ref{Details Appendix}. Results in subsequent appendices provide proofs of the results 
 given in the main text.
 
Proceeding, let $e$ and $q$ represent particular values (or realisations) of the random
variables $E$ and $Q$, respectively. In the analysis we present,
we will use the shorthand \textit{distribution} for a probability density
function, and, in particular, we will make extensive use of
\textit{conditional distributions}. For example, we will write $f_{Q|E}(q|e)$
for the conditional distribution of $Q$, when $E$ takes the value $e$ (i.e.,
when $E=e$). Thus, for the example of drying ground mentioned above, $E$ and
$Q$ would represent air temperature and drying time, respectively, while
$f_{Q|E}(q|e)$ would be the distribution of drying times when the air is at
the particular temperature $e$.

A natural way to characterise the relationship between $E$ and $Q$ is in terms of
information. That is, it is natural to ask: how much information does
knowledge of the value of $E$ provide about the value of $Q$ (and vice versa)? The
standard way to answer this question uses the idea of \textit{mutual information}, as
introduced by Claude Shannon in the mid 20th century (\cite{Shannon},
\cite{CoverThomas}). Mutual information is a powerful concept that has proved
to be extremely useful in both science and engineering (\cite{MacKay},
\cite{Brillinger}, \cite{Adesso}).

According to the concept of mutual information, when we receive information we
experience a decrease in `uncertainty.' Thus, if we know the actual
temperature in a room, then our uncertainty about the next reading we will
observe on a thermometer located in the room is decreased. Similarly, knowing
the thermometer's reading decreases our uncertainty about the air temperature.
Thus, the information is `mutual.'

Uncertainty is quantified by \textit{entropy} (\cite{Shannon},
\cite{CoverThomas}). For a one-dimensional (or scalar) random variable $X$,
with distribution $f_{X}(x)$, the entropy is $-\int_{-\infty}^{\infty
}f_{X}(x)\log_{2}\left[f_{X}(x)\right]  dx$ (\cite{Shannon},
\cite{CoverThomas}). Note that the range of the integral defining the entropy
need not be infinite\footnote{The limits of the $x$ integral
range from $-\infty$ to $\infty$, implicitly assuming that possible values of
$X$ lie in this infinite range. However, if $X$ takes values in a smaller
range, then $f_{X}(x)$ vanishes for $x$ outside this range (as does
$f_{X}(x)\log_{2}\left[  f_{X}(x)\right] $ when naturally defined as a limit).
Thus only $x$-values that are within the range of $X$ contribute to the
entropy.}.

Entropy is a measure of the extent a random variable is dispersed over its
range. For example, if there is a finite range of $X$ values, then the entropy
is maximised when $X$ is uniformly distributed over its range, and minimised
when its distribution is appreciable over only a very small range. This
justifies entropy as a measure of uncertainty.

While entropy is not identical to variance, various well-known families of
distributions (including Gaussian, uniform, exponential, and chi-squared) have
entropies that increase with variance, so large entropies are associated with
high variances.

We note that the entropy of a continuous random variable, as specified above
in the form of an integral, is often called \textit{differential entropy} in
the information-theoretic literature \cite{CoverThomas}. This phrase is used
to distinguish differential entropy from the entropy of a discrete random
variable. The distinction is important in some cases. However,
the focus of the current study is on mutual information, and in this context
the differences between differential entropy and the entropy of discrete
variables can largely be ignored. Perhaps for this reason, various authors use
the word \textquotedblleft entropy\textquotedblright\ to refer to both
differential entropy and to the entropy of discrete variables (\cite{Shannon},
\cite{CoverThomas}). We will follow this tradition here, and will use the term
`entropy' to refer to the differential entropy associated with
continuous variables.

Entropy is a characteristic of a probability distribution. When $E=e$ the
distribution of $Q$ is $f_{Q|E}(q|e)$, and the relevant entropy of $Q$ is that
associated with $f_{Q|E}(q|e)$, which we denote by $h(Q|e)$. We assume that
the value of $E$ varies from place to place (or that $E$ varies over time),
and write the distribution of $E$ as $f_{E}(e)$.

The (global) distribution of $Q$ values is obtained from an average over all
locations (or all times), and is denoted by $f_{Q}(q)$. It is given by
\begin{equation}
f_{Q}(q)=\int_{-\infty}^{\infty}f_{Q|E}(q|e)f_{E}(e)de.\label{f(q)}
\end{equation}
We can think of Eq. (\ref{f(q)}) as a specification of the distribution of $Q
$ that applies when the value of $E$ is not known. The entropy associated with
the distribution $f_{Q}(q)$ is written as $h(Q)$.

Let $I(Q; e)$ represent the information gained (or uncertainty decreased)
about the value of $Q$ when we observe that the environmental variable, $E$, takes
the value $e$ (that is, when $E = e$). From the definitions and considerations
given above we have
\begin{equation}
I(Q;e)=h(Q)-h(Q|e).\label{I(Q;e) def}
\end{equation}
The mutual information between $E$ and $Q$, denoted $I(Q;E)$,
corresponds to the \textit{average} amount of information gained about the
value of $Q$ when we observe the value of $E$. Thus $I(Q;E)$ is
the average over all $e$ of $I(Q;e)$, i.e.,
\begin{equation}
I(Q;E)=\int_{-\infty}^{\infty}I(Q;e)f_{E}(e)de.\label{I(Q;E) def}
\end{equation}
Note that $I(Q;E)$ is always non-negative (\cite{Shannon}, \cite{CoverThomas}).

In an entirely analogous way, we can condition on $Q=q$, leading
to the conditional distribution of $E$, which is given by $f_{E|Q}(e|q)$. We
then define: (i) the corresponding entropy of $E$ when $Q=q $, which we write
as $h(E|q)$, (ii) the entropy of $E$ when we \textit{do not} condition on the
value of $Q$, written $h(E)$, and associated with the distribution $f_{E}(e)$.

Using these definitions we can specify $I(E;q)$, the information that is
gained about the value of $E$ when $Q=q$:
\begin{equation}
I(E;q)=h(E)-h(E|q).\label{I(E;q) def}
\end{equation}
The average amount of information about $E$ that is obtained when the value of
$Q$ is observed is
\begin{equation}
I(E;Q)=\int_{-\infty}^{\infty}I(E;q)f_{Q}(q)dq.\label{I(E;Q) def}
\end{equation}
We are guaranteed that $I(E;Q)=I(Q;E)$ (\cite{Shannon}, \cite{CoverThomas}),
hence $I(Q;E)$ and $I(E;Q)$ are referred to as measures of \textit{mutual} information.

From Eqs. (\ref{f(q)}), (\ref{I(Q;e) def}) and (\ref{I(Q;E) def}) we can see
that mutual information depends on the distribution of $E$, as represented by
$f_{E}(e)$. Let $\tilde{f}_{E}(e)$ denote a distribution of $E$ that
\textit{maximises} the mutual information. Note that in what follows, we
shall indicate by a tilde,\quad$\widetilde{}$ , all quantities
that\textit{\ depend} on (or are) the mutual-information maximising
distribution of $E$.

Let $\widetilde{I}(E;Q)$ be the maximal value of $I(E;Q)$ (which is achieved
when the distribution of $E$ coincides with $\tilde{f}_{E}(e)$). The value of
$\widetilde{I}(E;Q)$ is known as the \textit{channel capacity}, and may be
thought of as a measure of how precisely the value of $E$ determines the value
of $Q$. (If there are multiple forms of $f_{E}(e)$ that all maximise the
mutual information, then we arbitrarily choose one of these and call it
$\tilde{f}_{E}(e)$.)

Let us now return to the matter of measurement. In general, there is no reason
to expect that the naturally occurring relationship between $E$ and $Q $ will
be similar to the engineered relationship between air temperature and the
measured temperature that would be apparent in a well-behaved thermometer,
such as the one described above. However, we can try to improve the situation
with the use of a transformation of $Q$ that creates a new variable with
desirable properties. To this end, we consider a transformation of $Q$ in the
form of a strictly increasing and differentiable function. We
call transformations of this sort \textit{increasing transformations}. As an
example, if $Q$ is a positive-valued variable that represents mass, then we
could transform $Q$ by taking the logarithm of mass. Like all increasing
transformations, this means there is a unique value of the transformed
variable (the logarithm of mass) for every possible value of the original
variable (mass). Furthermore, the transformed variable increases continuously
with $Q$, and is a differentiable function of $Q$. For additional properties
of increasing transformations, see Appendix \ref{Increasing Appendix}.

It can be shown that replacing $Q$ by an increasing transformation of $Q$ has
no effect on channel capacity (see \cite{Kraskov}, and for more details see
Appendix \ref{Increasing Appendix}). Thus, channel capacity is
invariant under all possible increasing transformations. Indeed, channel
capacity is shown in Appendix \ref{Increasing Appendix} to be invariant even
if we use two \textit{different} increasing transformations: one to transform
$Q$, the other to transform $E$. We will make use of this fact presently.

In order to determine one of the transformations we shall use, we introduce
\begin{equation}
\tilde{f}_{Q}(q)=\int_{-\infty}^{\infty}f_{Q|E}(q|e)\tilde{f}_{E}
(e)de\label{f tilde of q}
\end{equation}
which is the distribution of $Q$ that follows (from Eq. (\ref{f(q)})) when the
distribution of $E$ maximises the mutual information. For details of the
maximisation of the mutual information that determines $\tilde{f}_{E}(e)$, see
Appendix \ref{Maximum Appendix}.

Using $\tilde{f}_{E}(e)$ and $\tilde{f}_{Q}(q)$ we specify two increasing
transformations: one for $E$ and one for $Q$. We call the transformed
variables $E^{\ast}$ and $Q^{\ast}$, respectively. The transformation from $E
$ to $E^{\ast}$ can be specified in terms of the way a particular value of $E
$, say $e$, is transformed into the corresponding particular value of
$E^{\ast}$, which we write as $e^{\ast}$. The transformation is given by
\begin{equation}
e^{\ast}=\int_{-\infty}^{e}\tilde{f}_{E}(x)dx.\label{e* def}
\end{equation}
A shorter, equivalent way to define $E^{\ast}$ can be given\footnote{If we
define $\tilde{F}_{E}(e)=\int_{-\infty}^{e}\tilde{f}_{E}(x)dx$, then the
relation between $E$ and $E^{\ast}$ can be compactly written as $E^{\ast
}=\tilde{F}_{E}(E)$.}.

Similarly, the way a particular value of $Q$, say $q$, is transformed into the
corresponding particular value of $Q^{\ast}$, which we write as $q^{\ast} $,
is
\begin{equation}
q^{\ast}=\int_{-\infty}^{q}\tilde{f}_{Q}(x)dx.\label{q* def}
\end{equation}

Although the transformations in Eqs. (7) and (8) employ the
distribution of $E$ that maximises the mutual information ($\tilde{f}_{E}(e)$), 
we note that 
the random variable $E$ has the distribution $f_{E}(e)$, which generally differs from  $\tilde{f}_{E}(e)$.

In the case where the distribution of $E$ coincides with the
information-maximising distribution, $\tilde{f}_{E}(e)$, we denote the
resulting distributions of $E^{\ast}$ and $Q^{\ast}$ by $\tilde{f}_{E^{\ast}
}(e^{\ast})$ and $\tilde{f}_{Q^{\ast}}(q^{\ast})$, respectively.

Note that both $\tilde
{f}_{E^{\ast}}(e^{\ast})$ and $\tilde{f}_{Q^{\ast}}(q^{\ast})$ are
\textit{uniform} distributions on the interval $0$ to $1$. This is because
transformations of the form of Eqs. (\ref{e* def}) and (\ref{q* def}) produce
uniform distributions \cite{haigh2013probability}. The uniformity of $E^{\ast}$ and $Q^{\ast}$
is both convenient, and satisfyingly simple. In addition, uniform
distributions are what one might expect for a typical measuring instrument
that is functioning properly.

Let $f_{Q^{\ast}|E^{\ast}}(q^{\ast}|e^{\ast})$ represent the distribution of
$Q^{\ast}$ that applies when $E^{\ast}=e^{\ast}$. Thus, $f_{Q^{\ast}|E^{\ast}
}(q^{\ast}|e^{\ast})$ is the conditional distribution of the transformed
variable, $Q^{\ast}$. Let $h(Q^{\ast}|e^{\ast})$ represent the entropy
associated with $f_{Q^{\ast}|E^{\ast}}(q^{\ast}|e^{\ast})$.

We are now in a position to state our first theorem, which forms the primary
result presented in this work:


\begin{theorem}
\label{h=-Imax theorem} For all values of $e^{\ast}$ in the range $0<e^{\ast
}<1$ we have
\begin{equation}
h(Q^{\ast}|e^{\ast})=-\widetilde{I}(E;Q).\label{h=-Imax}
\end{equation}

\end{theorem}
\noindent Thus, the entropy of $Q^{\ast}$, when $E^{\ast}$ takes the particular value $e^{\ast}$ 
(i.e., when $E^{\ast}=e^{\ast}$), is entirely
independent of that particular value. In other words, the entropy, $h(Q^{\ast}|e^{\ast})$, 
always takes the same value, regardless of the value of the environmental variable, $e^{\ast}$. 
Furthermore, this \textit{universal value} of $h(Q^{\ast}|e^{\ast})$ is equal, in absolute value, 
to the channel capacity associated with the causal
relationship between $Q$ and $E$. We use the phrase \textit{homogenization of the entropy} to refer to the
independence of $h(Q^{\ast}|e^{\ast})$ from the value of $e^{\ast}$.

Note that, because $Q^{\ast}$ and $E^{\ast}$ are confined to a range between
zero and one, they can only have (differential) entropies that are less
than or equal to zero. Thus, Eq. (\ref{h=-Imax}) is consistent with the
fact that mutual information is always non-negative.

A proof of Theorem \ref{h=-Imax theorem} appears in Appendix
\ref{Special Appendix}.

As we shall see, $E^{\ast}$ has some additional convenient properties that $E
$ does not possess. However, we can also condition on the non-transformed
variable, $E$, and calculate $f_{Q^{\ast}|E}(q^{\ast}|e)$, the distribution of
$Q^{\ast}$ given that $E=e$. Doing this, we find that $h(Q^{\ast}|e)$, the
entropy of $f_{Q^{\ast}|E}(q^{\ast}|e)$, is independent of $e$, just as
$h(Q^{\ast}|e^{\ast})$ is independent of $e^{\ast}$. This is of interest
because it allows for a substantial generalisation of Theorem
\ref{h=-Imax theorem}. In particular, while we have assumed that $E$ is one
dimensional and continuous, this is not necessary for the
independence-of-entropy result embodied in Theorem \ref{h=-Imax theorem}. In
point of fact, $h(Q^{\ast}|e)$ is independent of the value of $e$ even if $E$
is a more general sort of random variable, for example discrete, or
multidimensional and continuous, or multidimensional with some dimensions
continuous, and others discrete (see Appendix \ref{Special Appendix} for more details.)

For many families of continuous probability distributions, the entropy is
completely determined by the variance, and vice versa. This is true, for
example, of uniform, Gaussian, exponential and chi-squared distributions. If
the conditional distributions are from a family of distributions with this
property, then $h(Q^{\ast}|e^{\ast})$ being independent of the value of
$e^{\ast}$ implies that the \textit{variance} of $f_{Q^{\ast}|E^{\ast}
}(q^{\ast}|e^{\ast})$ is also independent of the value of $e^{\ast}$.

As a consequence of Theorem \ref{h=-Imax theorem}, we know that, for any
choice of $f_{E}(e)$, the amount of information about the value of $Q^{\ast} $
that we obtain when we observe that $E^{\ast}=e^{\ast}$, namely $I(Q^{\ast
};e^{\ast})$, is the same for all possible values of $e^{\ast}$ (see (Eq.
(\ref{I(Q;e) def})). This is an encouraging and potentially useful result, as
one of our objectives is to ensure that measurement of the environment is
equally accurate and precise for all states of the environment. However, at
present there is no reason to expect, if $E^{\ast}=e^{\ast}$, that the values
of $Q^{\ast}$ that we obtain will tend to cluster around $e^{\ast}$, as we
would expect for a typical artificial measuring device. Another problem is
that the method we have described depends on knowing $\tilde{f}_{E}(e)$, which
is an information-maximising distribution of $E$. However, there is no general
method by which $\tilde{f}_{E}(e)$ can be calculated (though, in specific
situations, it can often be found or estimated \cite{CoverThomas},
\cite{Kraskov}). Finally, while $I(Q^{\ast};e^{\ast})$ is the same for all
values of $e^{\ast}$, the same is not generally true for $I(E^{\ast};q^{\ast
})$ (the information about the value of $E^{\ast}$ that we obtain when we
observe that $Q^{\ast}=q^{\ast}$). Thus, we may gain more information about
the value of $E^{\ast}$ when we observe certain values of $Q^{\ast}$, as
compared to other values of $Q^{\ast}$ (we will see an example of this below).
This does not suggest the sort of symmetry that we expect from a good
measuring device. We shall now show that all of these problems can be
ameliorated if we restrict the range of cases under consideration.

\subsection{Slow-change regime}

\indent Let us now consider a more restricted set of situations which
facilitate further analysis. We call this set of situations the
\textit{slow-change regime}. The slow-change regime is defined by a set of
additional assumptions. In particular, for the slow-change regime, we assume
that both $E$ and $Q$ have finite ranges, and that they are positively
correlated. For this last point, let $m(e)$ represent the mean value of $Q$
when $E=e$. That is, $m(e)$ is the mean of the conditional distribution
$f_{Q|E}(q|e)$. We incorporate positive correlation of $E$ and $Q$ by assuming
that $m(e)$ is a strictly increasing function of $e$. With the derivative of
$m(e)$ with respect to $e$ written $m^{\prime}(e)$, the property that $m(e)$
is strictly increasing corresponds to $m^{\prime}(e)>0$ for all allowed values
of $e$.

Let $\sigma_{\max}$ represent the maximum value of the standard deviation of
$f_{Q|E}(q|e)$, over all possible values of $e$. For the slow-change regime we
assume that $\sigma_{\max}$ is very small compared with the typical range over
which `shape-statistics' of $f_{Q|E}(q|e)$, such as the variance, the skew,
and the kurtosis, all appreciably change with $e$. We also assume that
$m^{\prime}(e)$ changes slowly in comparison to $\sigma_{\max}$. This is the
sense in which \textquotedblleft change\textquotedblright\ is
\textquotedblleft slow\textquotedblright\ in the slow-change regime.

The slow-change regime is relatively broad in the sense that the shape of
$f_{Q|E}(q|e)$ can be very different for two values of $e$ that differ by many
$\sigma_{\max}$. Furthermore, the relationship between the value of $e$ and
$m(e)$ can vary greatly with $e$. Thus, for example, $m(e)$ may increase
rapidly with $e$ when $e$ is small, and increase slowly with $e$ when $e$ is large.

We now give the form of the information-maximising distribution of $E$, namely
$\tilde{f}_{E}(e)$, under the assumptions of the slow-change regime. In
Appendix \ref{Slow Change Appendix} we prove the following theorem:


\begin{theorem}
\label{slow change theorem} Under the assumptions of the slow-change regime,
the value of $\tilde{f}_{E}(e)$ is proportional to
\begin{equation}
\frac{m^{\prime}(e)}{2^{h(Q|e)}}.\label{unnormalised f(e)}
\end{equation}
\end{theorem}


\noindent Thus, $\tilde{f}_{E}(e)$ is an increasing function of $m^{\prime
}(e)$, but a decreasing function of $h(Q|e)$.

A straightforward consequence of Theorem \ref{slow change theorem} is the
following corollary:


\begin{corollary}
Writing $e_{\min}$ and $e_{\max}$ for the minimum and maximum values,
respectively, that $E$ can take, the information-maximising distribution of
$E$ is given by
\begin{equation}
\tilde{f}_{E}(e)=\frac{1}{N} \times\frac{m^{\prime}(e)}{2^{h(Q|e)}
}\label{f tilde coroll}
\end{equation}
where
$N$ is the normalising factor $N=\int_{e_{\min}}^{e_{\max}} \frac{m^{\prime
}(x)}{2^{h(Q|x)}}dx$.
\end{corollary}


In Appendix \ref{Slow Change Appendix} we prove that the channel capacity
associated with a causal relationship is determined by the normalising factor,
$N$. In particular, we have the following additional corollary to Theorem
\ref{slow change theorem}:


\begin{corollary}
Under the assumptions of the slow-change regime, the channel capacity
associated with $f_{Q|E}(q|e)$ is given by
\begin{equation}
\widetilde{I}(Q;E)=\log_{2}\left(  \int_{e_{\min}}^{e_{\max}}\frac{m^{\prime}
(x)}{2^{h(Q|x)}}dx\right) .
\end{equation}
\end{corollary}


In Appendix \ref{Slow Change Appendix} we also show that, under the
assumptions of the slow-change regime, the relationship between the
transformed variables $E^{\ast}$ and $Q^{\ast}$ is similar to the relationship
between a well-made measuring device and the environmental variable it
measures. In particular, if we represent the mean value of $Q^{\ast}$ when
$E^{\ast}=e^{\ast}$ as $m^{\ast}(e^{\ast})$, then we have


\begin{corollary}
\label{m* corollory} Under the assumptions of the slow-change regime, when
$E^{\ast}=e^{\ast}$, the mean value of $Q^{\ast}$ is given by
\begin{equation}
m^{\ast}(e^{\ast})=e^{\ast}+O(\sigma_{max}).
\end{equation}
\end{corollary}


Corollary \ref{m* corollory} shows that our measuring variable is likely to
take values that cluster around the value of the environmental variable, as
would naturally be required of a good measuring device. Furthermore, recall
that: (i) for many commonly used continuous distributions, the entropy of the
distribution (i.e., $h(Q^{\ast}|e^{\ast})$) determines the variance of the
distribution (and vice versa), and (ii) $h(Q^{\ast}|e^{\ast})$ is independent
of the value of $e^{\ast}$. These two facts suggest that, typically, the
precision with which the value of $E^{\ast}$ predicts the value of $Q^{\ast}$
will be independent of the value of $E^{\ast}$. Thus, under the assumptions of
the slow-change regime, the relationship between our transformed variables
will tend to have the same character as that between a well-behaved measuring
device and the environmental variable it measures.

In addition to ensuring the highly suitable character of the mean and the
entropy associated with $f_{Q^{\ast}|E^{\ast}}(q^{\ast}|e^{\ast})$ , the
assumptions of the slow-change regime imply that our transformed variables
have other characteristics that one might expect from a good measuring device.
For example, as noted above, channel capacity is achieved when $E^{\ast}$ is
uniformly distributed. Under the slow-change regime, when $E^{\ast}$ takes its
uniform, information-maximising form, we can then calculate the conditional
distribution of $E^{\ast}$ given $Q^{\ast}=q^{\ast} $, namely $f_{E^{\ast
}|Q^{\ast}}(e^{\ast}|q^{\ast})$. In Appendix \ref{Slow Change Appendix} we
prove the following corollary to Theorem \ref{slow change theorem}:


\begin{corollary}
\label{he* corollory} Under the assumptions of the slow-change regime, when
$Q^{\ast}=q^{\ast}$ and when the distribution of $E^{\ast}$ is uniform, the
mean value of $E^{\ast}$ is given by:
\begin{equation}
q^{\ast}+O(\sigma_{max}).\label{mean E* corollory}
\end{equation}
Furthermore, under these conditions, $h(E^{\ast}|q^{\ast})$, which represents
the entropy of $E^{\ast}$ when $Q^{\ast}=q^{\ast}$, takes the same value for all allowable
values of $q^{\ast}$. This universal value of $h(E^{\ast}|q^{\ast})$ is given
by:
\begin{equation}
h(E^{\ast}|q^{\ast})=-I_{max}(E|Q).\label{hE*=-Imax corollory}
\end{equation}
\end{corollary}


Corollary \ref{he* corollory} implies that, under the slow-change regime, when
the distribution of $E^{\ast}$ is uniform, the values of $E^{\ast}$ that we
observe when $Q^{\ast}=q^{\ast}$ will tend to be close in value to $q^{\ast}$.
Furthermore, under these conditions, the information about the value of
$E^{\ast}$ that we obtain by observing the value of $Q^{\ast}$ is the same for
all possible values of $Q^{\ast}$. This is a consequence of Eq.
(\ref{hE*=-Imax corollory}), as one can see from Eq. (\ref{I(E;q) def}). Once
again, these results are consistent with what we might expect from a
well-behaved measuring device. Furthermore, the results suggest a certain
symmetry, in that the ability to predict the value of $Q^{\ast}$ by using the
value of $E^{\ast}$ is the same for all possible values of $E^{\ast}$, and
vice versa.

\subsection{Illustrative examples}

\indent In this section we present three examples in which the transformations
described above, in Eqs. (\ref{e* def}) and (\ref{q* def} ), have been
applied. The first two are calculated under the assumptions of the slow-change
regime, while the last example is not. These three examples illustrate three
different ways in which the transformations can operate to achieve their
effects. For ease of exposition, we have made some convenient choices about
the ranges of $E$ and $Q$.

Under the assumptions of the slow-change regime there are two sources of
statistical `noise' that can cause uncertainty about the state of the
environment (i.e., about the value of $E$) when a value of the measuring
variable ($Q$) is observed. The first of these sources of noise is a
relatively high level of variation in the value of $Q$ that may arise when the
environmental variable takes certain values. This source of noise relates to
the value of the denominator of the expression in Eq. (\ref{unnormalised f(e)}
), namely $2^{h(Q|e)}$, which increases in magnitude with the entropy of $Q$
associated with the particular value $e$ of $E$. An example of this sort of
noise is given in Fig. 1.

In Fig. 1, the mean value of $Q$ associated with any given value of $E$ is
equal to that given value, which is why the lines shown leading from $E$ to
$Q$ in Fig. 1a are vertical. However, the variation (i.e., the entropy) in $Q$
is larger when $E\geq1/2$, compared with when $E<1/2$. In this case, the
numerator of the expression in Eq. (\ref{unnormalised f(e)}), namely
$m^{\prime}(e)$, will equal unity for all values of $e$. However, the
denominator of this expression will be larger when $e\geq1/2$ than when
$e<1/2$. This leads to the form of $\tilde{f}_{E}(e)$ shown in Fig. 1b, and
thus (via Eq. (\ref{f(q)})) to the form of $\tilde{f}_{Q}(q)$ shown in Fig.
1c. When $Q^{\ast}$ is generated, this form of $\tilde{f}_{Q}(q)$ causes (via
Eq. (\ref{q* def})) a stretching-apart of $Q$ values for $Q<1/2$ and a
shrinking of the distance between $Q$ values for $Q\geq1/2$. This stretching
and shrinking equalises the entropy in $Q$ such that, for the transformed
variables ($E^{\ast}$ and $Q^{\ast}$), the entropy of $Q^{\ast}$ is
independent of the value of $E^{\ast}$. In this case, $E$ undergoes an
identical shrinking and stretching (via Eq. (\ref{e* def})), when $E^{\ast}$
is produced. This ensures that, in Fig. 1d, the mean value of the transformed
measured variable ($Q^{\ast}$) is approximately equal to the value of the
transformed environmental variable ($E^{\ast}$), just as was the case for the
untransformed variables.

\begin{figure}[tbh]
\centering
\includegraphics[height=12cm,width=8cm]{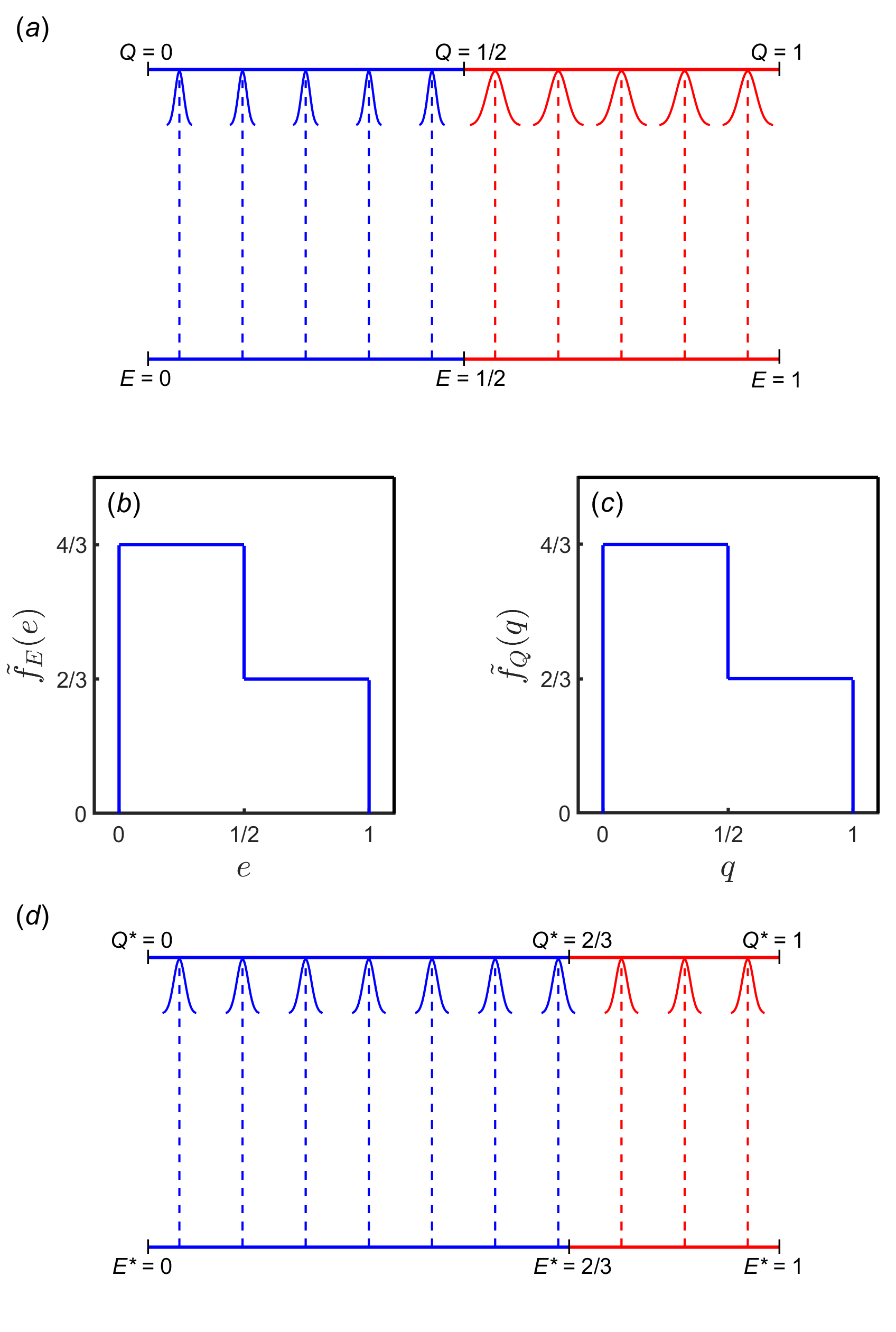}\caption{\textbf{Transformation
of variables when } $\boldsymbol{h(Q|e)}$\textbf{\ varies with }
$\boldsymbol{e}$\textbf{.} This figure is calculated under the approximation
of the slow-change regime. The figure applies for the special case $m(e)=e$,
where the conditional distribution $f_{Q|E}(q|e)$ takes two different forms,
depending on whether $e<1/2$ or $e\geq1/2$. The two forms of $f_{Q|E}(q|e)$
are such that the entropy, $h(Q|e)$, for $e\geq1/2$, exceeds the corresponding
entropy for $e<1/2$ by an amount of unity. \textbf{Panel (a):} This panel
contains dotted lines that connect values of $e$ with the associated values of
$m(e)$. The bell-shaped curves represent conditional distributions of $Q$
(i.e., $f_{Q|E}(q|e)$) centered on various values of $m(e)$. Note that, here
and in Fig. 2, these bell-shaped curves are illustrative, and are not drawn to
scale. \textbf{Panels (b\&c): }The marginal distributions $\tilde{f}_{E}(e)$
and $\tilde{f}_{Q}(q)$ are plotted for the situation described. \textbf{Panel
(d):} This panel is the analogue of Panel (\textbf{a}) of this figure, after
the transformations of Eqs. (\ref{e* def}) and (\ref{q* def}) have been
applied. Note that as a consequence of the transformations, the scale has been
stretched for $e<1/2$, and squeezed for $e\geq1/2$. For example, when $q=1/2$,
we have $q^{\ast}=2/3+O(\sigma_{\max})$. The same happens for $e$ and
$e^{\ast}$. As a result, for all values of $e^{\ast} $, the mean of
$f_{Q^{\ast}|E^{\ast}}(q^{\ast}|e^{\ast})$ equals $e^{\ast}+O(\sigma_{\max})$.
In addition, the entropy of $f_{Q^{\ast}|E^{\ast}}(q^{\ast}|e^{\ast})$ has the
same value for all values of $e^{\ast}$. In Panels \textbf{(a)} and
\textbf{(d)}, the parts shown in red relate to values of $e$ for which, before
transformation, $h(Q|e)$ is relatively large. }
\end{figure}


\begin{figure}[tbh]
\centering
\includegraphics[height=12cm,width=8cm]{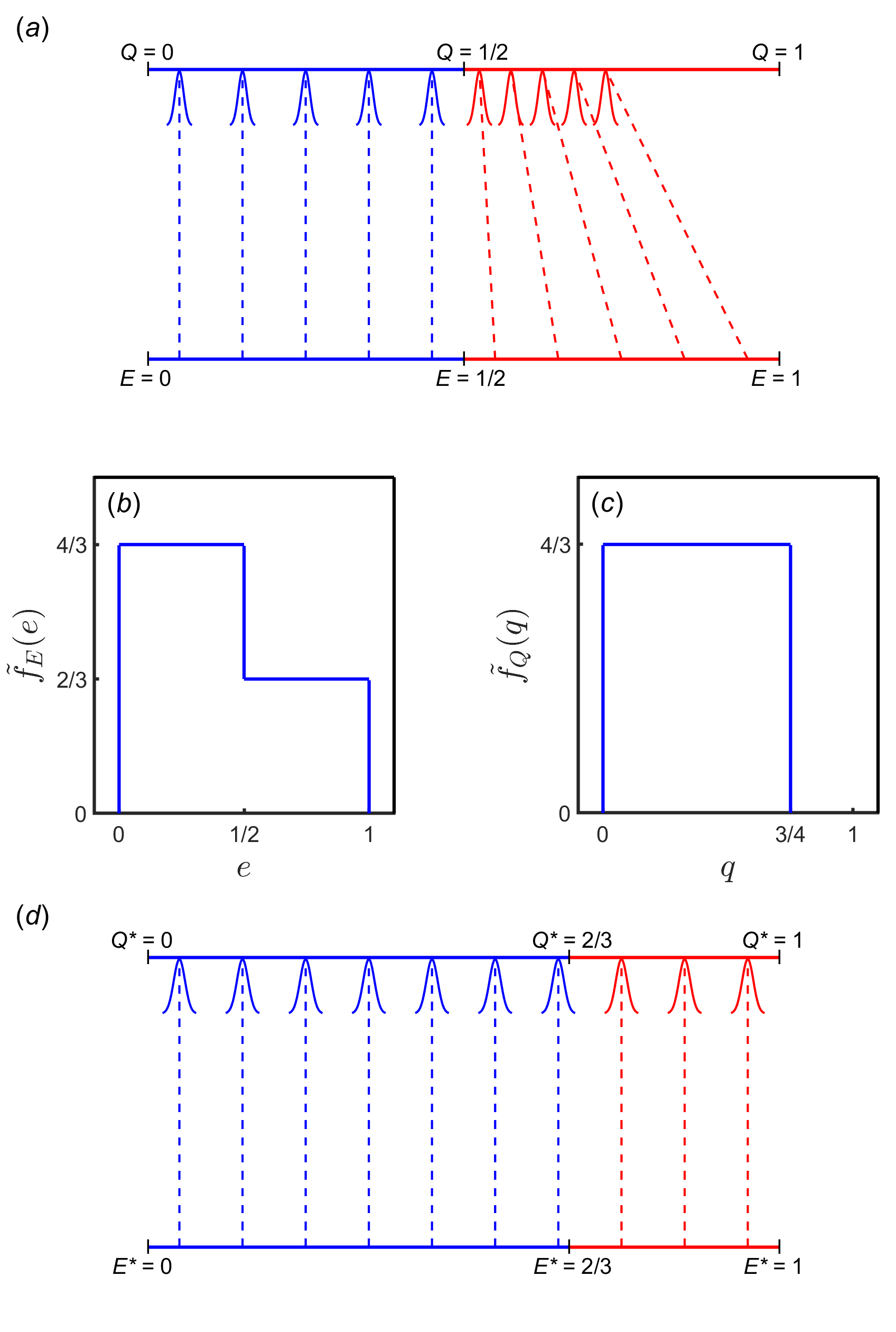}\caption{\textbf{Transformation
of variables when }$\boldsymbol{m} ^{\prime}\boldsymbol{(e)}$\textbf{\ varies
with }$\boldsymbol{e}$\textbf{.} This figure is calculated under the
approximation of the slow-change regime. The figure applies for the special
case where $m^{\prime}(e)=1$ for $e<1/2$, and $m^{\prime}(e)=1/2$ for
$e\geq1/2$. For this figure we assume that $h(Q|e)$ has the same value for all
possible values of $e$. \textbf{Panel (a): }This panel contains dotted lines
that connect values of $e$ with the associated values of $m(e)$.
\textbf{Panels (b\&c):}, The marginal distributions $\tilde{f}_{E}(e)$ and
$\tilde{f}_{Q}(q)$ are plotted for the situation described. \textbf{Panel
(d):} This panel is the analogue of Panel (\textbf{a}) of this figure, after
the transformations of Eqs. (\ref{e* def}) and (\ref{q* def}) have been
applied. The transformation from $E$ to $E^{\ast}$ is exactly as in Fig. 1.
The transformation, from $Q$ to $Q^{\ast} $, involves a uniform stretching of
the scale (by a factor of $4/3$) for $q<3/4$. Thus, when $q=1/2$ we have
$q^{\ast}=2/3+O(\sigma_{\max})$, and when $q=3/4$ we have $q^{\ast}
=1+O(\sigma_{\max})$. The differences in the way that $Q$ and $E$ are
transformed result in an alignment such that, for all values of $e^{\ast}$,
the mean of $f_{Q^{\ast}|E^{\ast}}(q^{\ast}|e^{\ast})$ is equal to $e^{\ast
}+O(\sigma_{\max})$. In Panels \textbf{(a)} and \textbf{(d)}, parts shown in
red relate to values of $e$ for which, before transformation, $m^{\prime
}(e)=1/2 $.}
\end{figure}


Fig. 2 relates to the second source of statistical noise that can cause
uncertainty about the state of the environment (i.e., about the value of $E$).
This source is associated with the numerator of Eq. (\ref{unnormalised f(e)}),
and it arises when there is overlap in the conditional distributions of $Q$
that are associated with two different values of $E$. This overlap, in turn,
depends on $m^{\prime}(e)$, the derivative, with respect to $e$, of the mean
of the conditional distribution of $Q$. Indeed, if $m^{\prime}(e)$ is a
constant for all values of $e$, then, for any two different values of $e $,
the separation of the means of the two associated distributions of $Q$
vanishes in the limit as $m^{\prime}(e)\rightarrow0$, and the overlap of the
two distributions approaches zero in the limit as $m^{\prime}(e)\rightarrow
\infty$.

In Fig. 2, the entropy associated with $f_{Q|E}(q|e)$ does not change as a
function of $e$. However, for $e<1/2$ we have $m^{\prime}(e)=1$, while for
$e\geq1/2$ we have $m^{\prime}(e)=1/2$. This leads to the form of $\tilde
{f}_{E}(e)$ shown in Fig. 2b, and thus to the form of $\tilde{f}_{Q}(q)$ shown
in Fig. 2c. In the construction of $E^{\ast}$, the form of $\tilde{f}_{E}(e)$
leads to exactly the same stretching and shrinking of the distance between
values of $E$ that was seen in Fig. 1. On the other hand, the form of
$\tilde{f}_{Q}(q)$ leads, in the construction $Q^{\ast}$, to a uniform
expansion of the space between values of $Q$. (This uniform expansion occurs
for all values of $Q$ that are associated with a non-zero probability
density.) As a result of these two different transformations, there is an
alignment of the values of $E^{\ast}$ with the means of the associated
conditional distributions of $Q^{\ast}$ values, as shown by the vertical
dotted lines in Fig. 2d. Furthermore, the uniform expansion that creates
$Q^{\ast}$ means that, for every possible environmental condition (i.e., for
every possible value of $e^{\ast}$), the entropy of $Q^{\ast}$ increases and
becomes equal (in absolute value) to the channel capacity.

\begin{figure}[tbh]
\centering
\includegraphics[height=12cm,width=8cm]{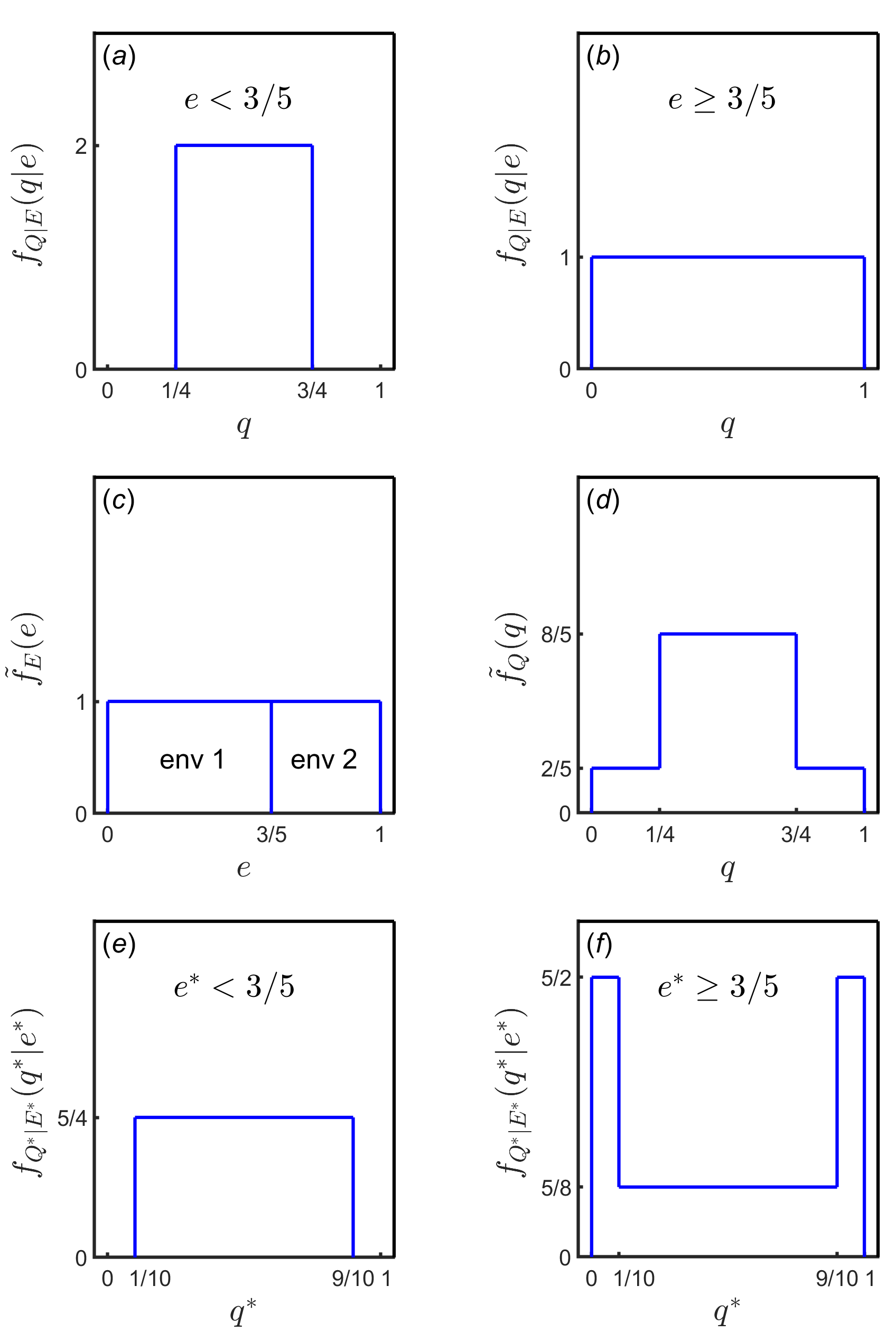}\caption{\textbf{Transformation
of variables when the variance cannot be homogenized.} In this case
$f_{Q|E}(q|e)$ is assumed to take two different forms. For $e<3/5$ the form of
$f_{Q|E}(q|e)$ is shown in \textbf{Panel (a)}, while for $e\geq3/5$ the form
of $f_{Q|E}(q|e)$ is shown in \textbf{Panel (b)}. \textbf{Panels (c\&d): }The
marginal distributions $\tilde{f}_{E}(e)$ and $\tilde{f}_{Q}(q)$ are plotted
for the situation described. \textbf{Panels (e\&f): }The two possible forms of
the conditional distribution of the measuring variable ($Q^{\ast}$), namely
$f_{Q^{\ast}|E^{\ast}}(q^{\ast}|e^{\ast})$, are plotted as functions of
$q^{\ast}$.}
\end{figure}


In Fig. 1 the transformation that constructs $Q^{\ast}$ ensures that
$h(Q^{\ast}|e^{\ast})$ is independent of $e^{\ast}$. This is accomplished by
stretching and shrinking the space between various values of $Q$ in a way that
may homogenize the variance of the conditional distributions of $Q^{\ast}$
that are associated with various values of $E^{\ast}$. However, it is of
interest to consider what happens when such a homogenization of variance is
not possible. A very simple example of such a situation is portrayed in Fig.
3. Here, we assume that only two forms of the conditional distribution of $Q$
are possible, and both forms have exactly the same mean. The first form of the
conditional distribution of $Q$ is associated with values of $E$ that are less
than $3/5$. This is a uniform distribution with mean equal to $1/2$ (see Fig.
3a). The second form of the conditional distribution of $Q$ is associated with
values of $E$ that equal or exceed $3/5$. This second form of the conditional
distribution also has a mean of $1/2$ and it is also uniform. However, its
variance is four times that of the first form of the distribution (see Fig.
3b). In Appendix \ref{Figure 3 Appendix} we show that no increasing
transformation (as defined above) can homogenize the variance such that, after
the transformation, the variance for the two possible forms of the conditional
distributions of the transformed measuring variable will be the same.

For this example, we picked $3/5$ as the critical value of $E$ because, when
$E$ is uniformly distributed from $0$ to $1$, this results in the
information-maximising distribution of $E$ (i.e., $\tilde{f}_{E}(e)$), which
is shown in Fig. 3c. From this we can obtain $\tilde{f}_{Q}(q)$, as shown in
Fig. 3d. Because $\tilde{f}_{E}(e)$ is uniform on the interval from $0$ to $1
$, application of Eq. (\ref{e* def}) has no effect, so that any distribution
of values of $E$ is identical to the associated distribution of values of
$E^{\ast}$ that is obtained from the transformation represented by Eq.
(\ref{e* def}).

In constructing $Q^{\ast}$, the form of $\tilde{f}_{Q}(q)$ leads to a uniform
expansion of the distance between values of $Q$ for those values of $Q$ that
lie between $1/4$ and $3/4$. For values of $Q$ that are above or below these
limits, the form of $\tilde{f}_{Q}(q)$ leads to a shrinking of the distance
between values of $Q$ when $Q^{\ast}$ is constructed. The result is the
creation of two possible forms for the conditional distribution of $Q^{\ast}$
(i.e., $f_{E^{\ast}|Q^{\ast}}(e^{\ast}|q^{\ast})$). The first of these is
associated with values of $e^{\ast}$ that are less than $3/5$, and this is
shown in Fig. 3e. The second form is for values of $e^{\ast}$ that equal or
exceed $3/5$, and is shown in Fig. 3f. Obviously, the variance of the
distribution shown in Fig. 3e is less than that for the distribution in Fig.
3f. Nevertheless, the entropy of these two distributions is identical, as
guaranteed by the analysis presented in Appendix \ref{Figure 3 Appendix}. This
Appendix also provides a calculation to demonstrate that, in this case, if
$E^{\ast}$ takes its uniform, information-maximising form, we nevertheless
gain more information about the value of $E^{\ast}$ when we observe some
values of $Q^{\ast}$ compared with other values. Note that, as mentioned
above, this cannot happen under the slow-change regime.

\section{Explaining the results}

\indent\ Explicit proofs of the claims made in this work are contained
in the Appendices. In this section we hope to give some insight into the
mathematical realities that lie behind the results, and into the logic
followed in the proofs that appear in the Appendices. We will focus on the
core result, as embodied in Theorem \ref{h=-Imax theorem}. This theorem states 
that the entropy of $Q^{\ast}$ is independent of the nature of the environment 
in which the entropy of $Q^{\ast}$ is measured, namely $h(Q^{\ast}|e^{\ast})$).

Note that this section is quite technical, and some readers may prefer to skip
directly to the Discussion, in which the possible wider implications of the
results are discussed.

\subsection{Why is the entropy of $Q^{\ast}$ independent of environmental
conditions?}

\indent It is reasonable to ask why the transformation specified by Eq.
(\ref{q* def}) leads to the homogenization of the entropy of the transformed
measuring variable ($Q^{\ast}$) that is found under different environmental
conditions. Here we will attempt, in a non-rigorous way, to provide some
insight as to why this result is true.

For our current purposes, it is convenient to define the \textit{conditional
entropy} of $Q$ as the weighted average of $h(Q|e)$ over the values of $e$
that occur. We write this weighted average as $h(Q|E)$ and thus have
$h(Q|E)=\int_{-\infty}^{\infty}h(Q|e)f_{E}(e)de$. This can be used to
rewrite Eq. (\ref{I(Q;E) def}) as:

\begin{equation}
I(Q;E)=h(Q)-h(Q|E).\label{mutual}
\end{equation}

Recall that the channel capacity is the maximum-possible value of $I(Q;E)$.
Channel capacity is achieved when the distribution of environmental effects
($f_{E}(e) $) takes on its information-maximising form, which we have called
$\tilde{f}_{E}(e)$. Now, let us imagine that, for sufficiently small
deviations in the distribution of environments from $\tilde{f}_{E}(e)$, the
value of $h(Q)$ (i.e., the entropy of $Q$) is unchanged. If this was true,
then what would it tell us about the entropy of $Q$ under various
environmental conditions? That is, what would it tell us about the value of
$h(Q|e)$ that we would find for various values of $e$?

The answer is that this odd set of circumstances would imply that there is no
variation among the values of $h(Q|e)$. To see why, imagine, for example, that
$\tilde{f}_{E}(e)$ is non-zero on the interval $(0,1)$, and that $h(Q|e) $ is
smaller, for all allowed $e<0.2$, compared with $h(Q|e)$ for all allowed
$e\geq0.2$. In this case, if $f_{E}(e)$ takes its information-maximising form
(i.e., $\tilde{f}_{E}(e)$), then we could decrease $h(Q|E)$ by increasing
$f_{E}(e)$ for $e<0.2$, while simultaneously decreasing $f_{E}(e)$ for
$e\geq0.2$. Note that, by our assumption, a sufficiently small change in
$f_{E}(e)$ from $\tilde{f}_{E}(e)$ will not alter $h(Q)$, and so this decrease
in $h(Q|E)$ (if the associated change in $f_{E}(e)$ is sufficiently small)
will lead to an increase in the mutual information, $I(Q;E)$ (Eq.
(\ref{mutual})). However, this increase in $I(Q;E) $ is impossible because it
arose by changing $f_{E}(e)$ from $\tilde{f}_{E}(e)$, and $\tilde{f}_{E}(e)$
is defined as the information-maximising form of $f_{E}(e)$. This logical
contradiction shows that our initial assumptions are incompatible. In other
words, if our assumption that $h(Q)$ is unchanged by sufficiently small
changes in $f_{E}(e)$ from $\tilde{f}_{E}(e)$ is true, then variation in
$h(Q|e)$ of the sort described must be impossible. By extending this example
to all possible cases of variation in the $h(Q|e)$ values, we can show that,
in general, the assumption about the immutability of $h(Q)$ when $f_{E}(e)$ is
sufficiently close to $\tilde{f}_{E}(e)$ implies that no variation is possible
among the values of $h(Q|e)$ that are associated with the various possible
values of $e $.

Of course, in general, $h(Q)$ will not be unaffected by changes in $f_{E}(e)$
when $f_{E}(e)$ is similar to $\tilde{f}_{E}(e)$, and thus variation in
$h(Q|e)$ is possible. However, consider what would happen if, when $f_{E}(e)$
takes its information-maximising form (namely $\tilde{f}_{E}(e)$), it so
happened that the resulting distribution of $Q$ (i.e., $\tilde{f}_{Q}(q)$) was
uniform on some interval (and had zero height elsewhere). 

As noted above, a uniform distribution maximises the entropy (in comparison with 
all other distributions that have non-zero probability density only on a 
particular interval). A direct implication of this maximisation is that very small 
changes in non-zero values of ${f}_{Q}(q)$ lead to yet smaller effects on $h(Q)$. 
In particular, if the deviation of  ${f}_{Q}(q)$ is of order $\alpha$, then 
the changes to $h(Q)$ will be of order $\alpha^{2}$.

Of course, changes in $h(Q)$ that are of order $\alpha^{2}$ are not the same
as no change at all to $h(Q)$. However, because $\alpha^{2}$ vanishes in
comparison to $\alpha$ as $\alpha$ approaches zero, it turns out that these
changes in $h(Q)$ are small enough to force equality among the values of
$h(Q|e)$ that obtain for different values of $e$. However, this equality
depends on our supposition that $\tilde{f}_{Q}(q)$ is uniform. In general, it
will not be uniform. However, we can use an increasing transformation to
create a new variable with a different distribution. A function that does this
in a way that ensures that the new distribution is uniformly distributed on
the interval $(0,1)$ is given by Eq. (\ref{q* def}). Thus, for the transformed
variable produced by Eq. (\ref{q* def}) (namely $Q^{\ast}$), it is plausible
that the entropy (namely $h(Q^{\ast}|e)$) must be equal for all possible
values of $e$.

\subsection{Summary of the proof that the entropy of $Q^{\ast}$ is
independent of environmental conditions}

\indent We shall now provide a sketch of the proof, contained in Appendices
\ref{Maximum Appendix} and \ref{Special Appendix}, that demonstrates the
independence of the entropy of $Q^{\ast}$ from the value of $e^{\ast}$ (which
is a measure of environmental conditions). These results arise, in part, from
the condition that the mutual information is maximised. To find this
condition, we look for the form of the distribution of $E$, such that to
linear order in changes in this distribution, the mutual information does not
change (in a functional sense, independence of changes, at linear order, is a
basic condition for stationarity). The form of the distribution of $E$ that
maximises the mutual information is written as $\tilde{f}_{E}(e)$. We find it
satisfies
\begin{equation}
\int_{-\infty}^{\infty}f_{Q|E}(q|e)\log_{2}\left[  \tilde{f}_{Q}(q)\right]
dq+h(Q|e)=-\widetilde{I}(Q;E)\label{condition maximum}
\end{equation}
with $\tilde{f}_{E}(e)$ implicitly present because of its control over the
form of $\tilde{f}_{Q}(q)$ (see Eq. (\ref{f tilde of q})). The freedom that
arises from the mutual information being unchanged in value, when the random
variables $Q$ and $E$ are replaced by new random variables that are increasing
transformations of $Q$ and $E$, allows us to transform the left side of Eq.
(\ref{condition maximum}), without changing the value of $\widetilde{I}(Q;E)$.
In particular, when the distribution of $E$ maximises the mutual information,
it is possible to find an increasing transformation that converts $Q$ to a new
random variable, $Q^{\ast}$, with the special property that its distribution,
$\tilde{f}_{Q^{\ast}}(q^{\ast})$, is uniform over $0<q^{\ast}<1$ and zero
elsewhere. When we use $Q^{\ast}$ in place of $Q$ in Eq.
(\ref{condition maximum}), we note that the first term on the left hand side
of this equation vanishes identically ($\log_{2}\left[  1\right]  =0$) and we
arrive at $h(Q^{\ast}|e)=-\widetilde{I}(Q;E)$. This result indicates that the
entropy $h(Q^{\ast}|e)$ has been homogenized in the sense that it is
independent of the value of $e$. Additionally, we are still free to transform
$E$. It is also possible to find an increasing transformation that converts
$E$ to a new random variable $E^{\ast}$ that is uniform over $0<e^{\ast}<1$
and zero elsewhere. This leads to the result $h(Q^{\ast}|e^{\ast})=-I_{\max
}(Q;E)$ and indicates that the entropy $h(Q^{\ast}|e^{\ast})$ is independent
of $e^{\ast}$ for $0<e^{\ast}<1$.

\section{Discussion}

\indent As we have seen, transformations of the sort described by Eqs.
(\ref{e* def}) and (\ref{q* def}) do not change the amount of information that
knowledge about one variable provides about the state of another variable.
With that in mind, it is worth considering whether such transformations can be
of any practical use.

Under certain conditions, the transformations described by Eqs. (\ref{e* def})
and (\ref{q* def}) make natural relationships behave in a manner that is
similar to the relationship between a continuously varying environmental
variable and a typical measuring device that is behaving properly. The fact
that humans go to so much trouble to produce good measuring devices suggests
that, in itself, this effect of the transformations is valuable. They bring a
certain homogeneity in that, after they are applied, the effect of the
environmental variable on the measuring variable can be about the same
throughout the range of possible environmental-variable values. Humans
apparently find this sort of homogeneity to be pleasing and/or useful.

A very practical possible application of the results presented here has to do
with statistical analysis. For data sets that arise from causal relationships
that are sufficiently similar to the situation described by the slow-change
regime, the methods described here can transform the data so as to linearise
the relationship between two data sets, and to homogenize the entropy among
samples collected under different conditions. As entropy is often closely
related to variance, this may imply a homogenization of variance as well.
Regression analysis and similar techniques often assume a linear relationship
between independent and dependent variables, and they also typically assume
that the variance is homogeneous. Thus, the transformations described here may
be useful in adjusting data so as to meet the requirements for the most
powerful statistical techniques available for data analysis \cite{Casella}.
Even for situations that are very different from those described by the
slow-change regime, a transformation of the sort described here can typically 
be used to homogenize the entropy of $Q$, the measuring variable, and it seems 
likely that this will be advantageous when statistical analyses are carried out.

Given that the transformations described here homogenize entropy in measuring
variables (which may or may not have implications for the homogenization of
the variance) it is interesting to speculate on the possibility of developing
hypothesis-testing analyses in which the measure of data dispersion is
entropy, and not the variance (\cite{Vasicek}, \cite{Chen}, \cite{Baran}).
This possibility is particularly intriguing because there are situations in
which entropy can be homogenized, but variance cannot. (A simple example is
provided in Fig. 3.) In this regard, the transformations bear some resemblance 
to \textit{histogram equalisation}, which is a method employed to enhance low-quality optical images, 
amongst other uses (\cite{Laughlin}, \cite{Tkacik1}, \cite{Tkacik2},
\cite{Gonzalez}). However, this resemblance is in appearance only, since histogram
equalisation does not generally lead to a homogenization of entropy between environmental 
conditions.  This is due, in part, to the fact that, unlike the transformations we have 
described in the present study, histogram equalisation does not involve using the information-
maximising distribution of the causal variable to create a transformed version of the caused 
variable.  Instead, in histogram equalisation, the distribution of inputs (the causal 
variable) is typically taken from real-world data.  For example, to use histogram equalisation 
to enhance a digital image of a countryside scene, one might use the distribution of 
brightness levels that are reflected from objects in a natural landscape.

\bigskip

In this work we have focused on the information-theoretic
analysis of continuous variables. However, discrete variables have generally
attracted much more attention from information theorists than have continuous
variables. It is not obvious how to apply some of the concepts that have been
developed in the realm of discrete variables to the case of continuous
variables. One example of this difficulty has to do with the concept of
\textit{functional information}, as proposed by J. Szostak (\cite{Szostak},
\cite{Hazen}). Szostak, writing in the context of biomolecules such as
enzymes, says that \textquotedblleft functional information is simply
$-\log_{2}$ of the probability that a random sequence will encode a molecule
with greater than any given degree of function\textquotedblright
\ \cite{Szostak}. In addition to its use in the context of biomolecules,
functional information (or a very similar concept) has also been used to
characterise adaptation \cite{PeckWaxman2018} and the closely allied concept
of `biological complexity' (\cite{Adami2000}, \cite{Adami2002}). In these
cases, expected reproductive success (i.e., fitness) is the `function' in
terms of which the functional information associated with different types of
organisms is evaluated. However, these wider applications have been confined
to studying the adaptedness of genomes (or the biological complexity of
genomes). This is problematic because it is in the realm of phenotypes that
adaptedness is generally recognised. We infer a highly adapted genotype when
we see a highly adapted phenotype, and, in general, not vice versa. Thus, it
would be advantageous to be able to apply the idea of functional information
to the continuously varying traits that are typically used to characterise
phenotypes.

Unfortunately, the value of functional information for a continuously varying
phenotypic variable will, in general, depend on how that variable is
transformed. For example, if the variable is body mass, then we might measure
the fitness associated with different body-mass values (holding all else
constant), and thus we could calculate the functional information associated
with any given body mass. However, we may get very different values for
functional information if, instead of body mass, we consider the logarithm of
body mass. This would introduce problematic ambiguity if there was no natural
transformed variable with which to measure body mass. However, the variation
in body mass that we see within groups of organisms tends to relate, in part,
to the different evolutionary pressures that prevail in various environments
(\cite{Brown}, \cite{Alroy}, \cite{Morand}). This relationship might usefully
be characterised as a causal relationship of the sort discussed above. As
such, the results presented here suggest that, typically, there will be a
natural transformed variable to use in the measurement of body mass (or
whatever other continuously varying trait is being considered). This, in turn,
suggests that the results presented here may facilitate the extension of
information-theoretic concepts that were developed in the context of discrete
variables to the realm of continuously distributed variables. The implications
of the resulting increase in analytic power are not clear, but they may
include the development of useful quantitative tools to study some of the most
fascinating phenomena that are associated with life.\bigskip

{\large {\textbf{Acknowledgements}} }\bigskip

It is a pleasure to thank Antonio Carvajal Rodriguez, Yuval Simons, 
and John Welch for helpful discussions during the preparation of this study,
and Professor C. Adami, Professor O. Johnson and an anonymous reviewer for 
very helpful comments on our manuscript.

\newpage\appendices

\begin{center}
{\large {\textbf{APPENDICES}} }
\end{center}

In the following appendices, when we refer to a particular
\textit{distribution} we mean a particular `probability density function', and
when we refer to an \textit{entropy}, we mean the `differential entropy'
(which is the entropy associated with a continuous random variable
\cite{CoverThomas}).

We shall also make use of the Dirac delta function which, for argument $x$, is
written $\delta(x)$. The Dirac delta function, $\delta(x)$, is a spike-like
probability density, with a vanishingly small variance and an area of unity
that is located at $x=0$. We shall freely exploit the following two properties
of a Dirac delta function: (i) with $\mathbb{E}\left[  ...\right]  $ denoting
an expected value, the joint probability density function of two random
variables $E$ and $Q$, when evaluated at the values $e$ and $q$, respectively,
is $\mathbb{E}\left[  \delta\left(  e-E\right)  \delta\left(  q-Q\right)
\right]  $ (see e.g., the textbook \cite{van_Kampen} for this point of view);
(ii) when a function of $x$, say $q(x)$, vanishes at only one point, say
$x_{0}$, the quantity $\delta(q(x))$ equals $\delta(x-x_{0})/|q^{\prime}
(x_{0})|$ where $q^{\prime}(x)=dq(x)/dx$ (see e.g., \cite{Barton}).


\section{Mathematical details of the model}
\label{Details Appendix}


In this appendix we introduce the form of the joint distribution of the
random variables $Q$ and $E$ that we consider in this work. The joint
distribution of $Q$ and $E$ arises from a fixed form of the conditional
distribution, $f_{Q|E}(q|e)$, but different forms of the marginal distribution
of $E$, namely $f_{E}(e)$.

This appendix also contains basic mathematical details of the model adopted,
along with definitions of some key distributions and the mutual information.

To begin, consider a mathematical model involving two continuous
one-dimensional random variables that we write as $E$ and $Q$. Generally, $E$
and $Q$ are not statistically independent.

We shall write the limits of various integrals involving distributions of the
random variables $E$ and $Q$ as ranging from $-\infty$ to $\infty$, implicitly
assuming that possible values of $E$ and $Q$ lie in this infinite range.
However, when we consider new random variables (related to $E$ and $Q$) that
take values in a smaller range, the corresponding distributions will vanish
outside the ranges of the new variables, and make no contribution to the
integrals. When we need to be explicit about finite ranges of random
variables, we will indicate this in the limits of any integrals that arise.

\subsection{\label{Distributions}Distributions}

Some properties of important distributions are as follows.

\begin{enumerate}
\item The conditional distribution of $Q$, given that $E$ takes the value $e$
(i.e., given that $E=e$), is written as $f_{Q|E}(q|e)$. This, like all
probability density functions, has a total integrated probability of unity,
and for the present case this reads
\begin{equation}
\int_{-\infty}^{\infty}f_{Q|E}(q|e)dq=1.\label{normalisation f Q|E}
\end{equation}

A key assumption of this work is that $f_{Q|E}(q|e)$ has its $q$ and $e$
dependence \textit{specified at the outset}, and its specified form is
\textit{not varied} in the ensuing analysis. By contrast, we shall consider
different forms of the \textit{marginal distribution} of $E$, written $f_{E}(e)$.

\item The joint distribution of $Q$ and $E$ is given in terms of
$f_{Q|E}(q|e)$ and $f_{E}(e)$ as
\begin{equation}
f_{Q,E}(q,e)=f_{Q|E}(q|e)f_{E}(e).\label{f Q E}
\end{equation}

\item The marginal distribution of $Q$ is given by
\begin{equation}
f_{Q}(q)=\int_{-\infty}^{\infty}f_{Q|E}(q|e)f_{E}(e)de.\label{f Q}
\end{equation}

\end{enumerate}

We note that because of the fixed nature of $f_{Q|E}(q|e)$, different forms of
$f_{E}(e)$ produce different statistical properties of both $E$ and $Q$. In
particular, Eqs. (\ref{f Q E}) and (\ref{f Q}) explicitly show that the
distributions $f_{Q,E}(q,e)$ and $f_{Q}(q)$ depend on the marginal
distribution of $E$, namely $f_{E}(e)$. As a consequence, a change in
$f_{E}(e)$ induces changes in both $f_{Q,E}(q,e)$ and $f_{Q}(q)$.

\subsection{Differential entropy and mutual information}

A continuous random variable, such as $Q$, has an entropy (strictly,
\textit{differential} entropy) that we denote by $h(Q)$, and is defined by
\begin{equation}
h(Q)=-\int_{-\infty}^{\infty}f_{Q}(q)\log_{2}\left[  f_{Q}(q)\right]
dq\label{h(Q)}
\end{equation}
where $\log_{2}\left(  x\right)  $ denotes the logarithm of $x$ to base $2$.

Let us suppose that the random variable $E$ is observed to take the particular
value $e$ (i.e., $E=e$). Then the relevant distribution of $Q$ is the
conditional probability density of $Q$, given that $E=e$, namely
$f_{Q|E}(q|e)$. The entropy associated with $Q$ in this case is written as
$h(Q|e)$, and is calculated from $f_{Q|E}(q|e)$ according to
\begin{equation}
h(Q|e)=-\int_{-\infty}^{\infty}f_{Q|E}(q|e)\log_{2}\left[  f_{Q|E}
(q|e)\right]  dq.\label{h(Q|e)}
\end{equation}

The \textit{mutual information} is defined as
\begin{align}
I(Q;E)  & =\int_{-\infty}^{\infty}\left[  h(Q)-h(Q|e)\right]  f_{E}%
(e)de\nonumber\\
& \nonumber\\
& =h(Q)-\int_{-\infty}^{\infty}h(Q|e)f_{E}(e)de.\label{mutual I}%
\end{align}
This is closely analogous to the mutual information of a pair of discrete
random variables, which corresponds to the average reduction in the
uncertainty of $Q$ that results from knowledge of the value of $E$.


\section{Increasing transformations}
\label{Increasing Appendix}


In this appendix we demonstrate that the value of the mutual information is
unchanged on replacing the random variables $Q$ and $E$ by independent (and
generally different) transformations of these random variables. The
transformations are implemented with functions that are at least once
differentiable and strictly increasing and in this work are termed
\textit{increasing transformations}.

In Appendix \ref{Details Appendix}, we started with the random variables $Q$
and $E$. We now consider \textit{transformed versions} of these variables.
Using transformed versions of $Q$ and $E$ gives us the freedom to make
different choices of the transformations adopted.

To define the transformed variables, we introduce two real functions, $P(x)$
and $D(x)$, that are at least once differentiable and are \textit{strictly
increasing}. We shall call such functions `increasing transformations'. We
then define a pair of new continuous random variables, $Q^{\circ}$ and
$E^{\circ}$ via
\begin{equation}
Q^{\circ}=P(Q)\label{P}
\end{equation}
and
\begin{equation}
E^{\circ}=D(E)\label{D}
\end{equation}
which are thus increasing transformations of the random variables $Q$ and $E$,
respectively. Note that Eq. (\ref{P}) is equivalent to $q^{\circ}=P(q)$, where
$q$ is a particular value of $Q$ and $q^{\circ}$ is the corresponding
particular value of $Q^{\circ}$ that arises. Similarly, Eq. (\ref{D}) is
equivalent to $e^{\circ}=P(e)$, where $e$ is a particular value of $E$ and
$e^{\circ}$ is the corresponding particular value of $E^{\circ}$ that arises.

Increasing transformations are invertible, which means, for example, that
$Q^{\circ}$ uniquely determines $Q$ and vice versa. Thus Eqs. (\ref{P}) and
(\ref{D}) can also be written as $Q=P^{(-1)}(Q^{\circ})$ and $E=D^{(-1)}
(E^{\circ})$ where a $(-1)$ superscript denotes the \textit{inverse function},
such that $P(P^{(-1)}(x))=x$ and $P^{(-1)}(P(x))=x$.

It might be expected that $Q^{\circ}$ and $E^{\circ}$ are, in some sense, an
equivalent way of describing the problem at hand. Indeed, we shall show that
the mutual information between $Q$ and $E$, is identical to the mutual
information between $Q^{\circ}$ and $E^{\circ}$. Thus, at the level of mutual
information, the transformed pair of variables, $Q^{\circ}$ and $E^{\circ}$,
are \textit{completely equivalent} to the original pair of variables, $Q$ and
$E$. Of course some versions of $Q^{\circ}$ and $E^{\circ} $ may have some
additional properties that make them more useful than others.

We next show how some key statistical properties of the transformed variables
($Q^{\circ}$ and $E^{\circ}$) are related to the corresponding properties of
the original variables ($Q$ and $E$). Since, in general, $Q^{\circ}$ and
$E^{\circ}$ do not take the same range of values as the original variables, we
shall incorporate this into the analysis by taking $Q^{\circ}$ to lie in the
range $q_{1}^{\circ}$ to $q_{2}^{\circ}$, and $E^{\circ}$ to lie in the range
$e_{1}^{\circ}$ to $e_{2}^{\circ}$.

\subsection{Conditional distributions $f_{Q^{\circ}|E^{\circ}}(q^{\circ
}|e^{\circ})$ and $f_{E^{\circ}|Q^{\circ}}(e^{\circ}|q^{\circ})$}

We stated in Appendix \ref{Details Appendix} that the conditional distribution
$f_{Q|E}(q|e)$ is specified from the outset. It is convenient, however, to
represent it in a form where it can be related to the corresponding
distribution involving the transformed variables $Q^{\circ}$ and $E^{\circ}$.
With $\mathbb{E}\left[  ...\right]  $ denoting an expected value over $Q$ and
$E$, and $\delta(x)$ denoting a Dirac delta function of argument $x$, the
required representation of $f_{Q|E}(q|e)$ is given by
\begin{equation}
f_{Q|E}(q|e)=\frac{\mathbb{E}\left[  \delta\left(  q-Q\right)  \delta
(e-E)\right]  }{\mathbb{E}\left[  \delta(e-E)\right]  }.\label{fQ|E rep}
\end{equation}
Consider now the corresponding conditional distribution of $Q^{\circ}$,
conditional on the value of $E^{\circ}$, when evaluated at $q^{\circ}$ and
$e^{\circ}$, respectively, namely $f_{Q^{\circ}|E^{\circ}}(q^{\circ}|e^{\circ
})$. This can be similarly written as
\begin{align}
f_{Q^{\circ}|E^{\circ}}(q^{\circ}|e^{\circ})  & =\frac{\mathbb{E}\left[
\delta\left(  q^{\circ}-Q^{\circ}\right)  \delta(e^{\circ}-E^{\circ}\right]
}{\mathbb{E}\left[  \delta(e^{\circ}-E^{\circ}\right]  }\nonumber\\
& \nonumber\\
& =\frac{\mathbb{E}\left[  \delta\left(  q^{\circ}-P(Q)\right)  \delta
(e^{\circ}-D(E))\right]  }{\mathbb{E}\left[  \delta(e^{\circ}-D(E))\right]  }.
\end{align}

This is well defined for $e_{1}^{\circ}<e^{\circ}<e_{2}^{\circ}$ and in this
range we have
\begin{align}
& f_{Q^{\circ}|E^{\circ}}(q^{\circ}|e^{\circ})\nonumber\\
& \nonumber\\
& =\left\{
\begin{array}
[c]{ll}%
\tfrac{\mathbb{E}\left[  \delta\left(  P^{(-1)}(q^{\circ})-Q\right)
\delta(D^{(-1)}(e^{\circ})-E)\right]  }{P^{\prime}(P^{(-1)}(q^{\circ
}))\mathbb{E}\left[  \delta(D^{(-1)}(e^{\circ})-E)\right]  } & \text{for
}q_{1}^{\circ}<q^{\circ}<q_{2}^{\circ}\\
& \\
0 & \text{otherwise.}%
\end{array}
\right.  \nonumber\\
&
\end{align}
This can then be directly expressed in terms of $f_{Q|E}(q|e)$, using Eq.
(\ref{fQ|E rep}), as
\begin{align}
& f_{Q^{\circ}|E^{\circ}}(q^{\circ}|e^{\circ})\nonumber\\
& \nonumber\\
& =\left\{
\begin{array}
[c]{lll}%
\tfrac{f_{Q|E}(P^{(-1)}(q^{\circ})|D^{(-1)}(e^{\circ}))}{P^{\prime}%
(P^{(-1)}(q^{\circ}))} &  & \text{for }q_{1}^{\circ}<q^{\circ}<q_{2}^{\circ}\\
&  & \\
0 &  & \text{otherwise.}%
\end{array}
\right.  \nonumber\\
& \label{f Qdot|Edot}%
\end{align}

In a similar way the distribution $f_{E^{\circ}|Q^{\circ}}(e^{\circ}|q^{\circ
})$ is well defined for $q_{1}^{\circ}<q^{\circ}<q_{2}^{\circ}$ and in this
range is given by
\begin{align}
& f_{E^{\circ}|Q^{\circ}}(e^{\circ}|q^{\circ})\nonumber\\
& \nonumber\\
& =\left\{
\begin{array}
[c]{lll}%
\tfrac{f_{E|Q}(D^{(-1)}(e^{\circ})|P^{(-1)}(q^{\circ}))}{D^{\prime}%
(D^{(-1)}(e^{\circ}))} &  & \text{for }e_{1}^{\circ}<e^{\circ}<e_{2}^{\circ}\\
&  & \\
0 &  & \text{otherwise.}%
\end{array}
\right.  \nonumber\\
& \label{f Edot|Qdot}%
\end{align}

\subsection{Marginal distributions $f_{E^{\circ}}(e^{\circ})$ and
$f_{Q^{\circ}}(q^{\circ})$}

The marginal distribution of $E$, when evaluated at $e$, can be written as
$f_{E}(e)=\mathbb{E}\left[  \delta(e-E)\right]  $. The marginal distribution
of $E^{\circ}$, when evaluated at $e^{\circ}$, is given by
\begin{equation}
f_{E^{\circ}}(e^{\circ})=\mathbb{E}\left[  \delta(e^{\circ}-E^{\circ})\right]
=\mathbb{E}\left[  \delta(e^{\circ}-D(E))\right]  .
\end{equation}
Since $E^{\circ}$ only takes values in the range $e_{1}^{\circ}$ to
$e_{2}^{\circ}$ we have
\begin{equation}
f_{E^{\circ}}(e^{\circ})=\left\{
\begin{array}
[c]{lll}
\dfrac{\mathbb{E}\left[  \delta(D^{(-1)}(e^{\circ})-E)\right]  }{D^{\prime
}(D^{(-1)}(e^{\circ}))} &  & \text{for }e_{1}^{\circ}<e^{\circ}<e_{2}^{\circ
}\\
&  & \\
0 &  & \text{otherwise.}
\end{array}
\right.
\end{equation}
This can be written as
\begin{equation}
f_{E^{\circ}}(e^{\circ})=\left\{
\begin{array}
[c]{lll}
\dfrac{f_{E}(D^{(-1)}(e^{\circ})))}{D^{\prime}(D^{(-1)}(e^{\circ}))} &  &
\text{for }e_{1}^{\circ}<e^{\circ}<e_{2}^{\circ}\\
&  & \\
0 &  & \text{otherwise.}
\end{array}
\right. \label{marginal Edot}
\end{equation}
Similarly, the marginal distribution of $Q^{\circ}$, when evaluated at
$q^{\circ}$, is given by
\begin{equation}
f_{Q^{\circ}}(q^{\circ})=\left\{
\begin{array}
[c]{lll}
\dfrac{f_{Q}(P^{(-1)}(q^{\circ}))}{P^{\prime}(P^{(-1)}(q^{\circ}))} &  &
\text{for }q_{1}^{\circ}<q^{\circ}<q_{2}^{\circ}\\
&  & \\
0 &  & \text{otherwise.}
\end{array}
\right. \label{marginal Qdot}
\end{equation}

\subsection{Differential entropy $h(Q^{\circ})$}

We write the (differential) entropy of $Q$ as $h(Q)$. This is given in Eq.
(\ref{h(Q)}). The corresponding entropy of $Q^{\circ}$ is written as
$h(Q^{\circ})$ and given by $h(Q^{\circ})=-\int_{q_{1}^{\circ}}^{q_{2}^{\circ
}}f_{Q^{\circ}}(q^{\circ})\log_{2}\left[  f_{Q^{\circ}}(q^{\circ})\right]
dq^{\circ}$. Using Eq. (\ref{marginal Qdot}) yields\newline$h(Q^{\circ}
)=-\int_{q_{1}^{\circ}}^{q_{2}^{\circ}}\frac{f_{Q}(P^{(-1)} (q^{\circ}
))}{P^{\prime}(P^{(-1)}(q^{\circ}))}\log_{2}\left[  \frac{f_{Q}(P^{(-1)}
(q^{\circ}))}{P^{\prime}(P^{(-1)}(q^{\circ}))}\right]  dq^{\circ}$ and with
the change of variables $q^{\circ}=P(q)$ we obtain
\begin{align}
h(Q^{\circ})  & =-\int_{-\infty}^{\infty}f_{Q}(q)\log_{2}\left[  \frac
{f_{Q}(q)}{P^{\prime}(q)}\right]  dq\nonumber\\
& \nonumber\\
& =h(Q)+\int_{-\infty}^{\infty}f_{Q}(q)\log_{2}\left[  P^{\prime}(q)\right]
dq.
\end{align}

\subsection{Differential entropy $h(Q^{\circ}|e^{\circ})$}

We write the entropy of $Q$, when $E$ takes the particular value $e$, as
$h(Q|e)$. This is given in Eq. (\ref{h(Q|e)}). We write the corresponding
entropy of $Q^{\circ}$, given that $E^{\circ}$ takes the particular value
$e^{\circ}$, as $h(Q^{\circ}|e^{\circ})$ and this is given by
\begin{align}
h(Q^{\circ}|e^{\circ}) &  =-\int_{q_{1}^{\circ}}^{q_{2}^{\circ}}f_{Q^{\circ
}|E^{\circ}}(q^{\circ}|e^{\circ})\log_{2}\left[  f_{Q^{\circ}|E^{\circ}%
}(q^{\circ}|e^{\circ})\right]  dq^{\circ}\nonumber\\
& \nonumber\\
&  =-\int_{q_{1}^{\circ}}^{q_{2}^{\circ}}\frac{f_{Q|E}(P^{(-1)}(q^{\circ
})|D^{(-1)}(e^{\circ}))}{P^{\prime}(P^{(-1)}(q^{\circ}))}\nonumber\\
& \nonumber\\
&  \quad\times\log_{2}\left[  \frac{f_{Q|E}(P^{(-1)}(q^{\circ})|D^{(-1)}%
(e^{\circ}))}{P^{\prime}(P^{(-1)}(q^{\circ}))}\right]  dq^{\circ}\nonumber\\
& \nonumber\\
&  =-\int_{-\infty}^{\infty}f_{Q|E}(q|D^{(-1)}(e^{\circ}))\nonumber\\
& \nonumber\\
&  \quad\times\log_{2}\left[  \frac{f_{Q|E}(q|D^{(-1)}(e^{\circ}))}{P^{\prime
}(q)}\right]  dq\nonumber\\
& \nonumber\\
&  =h(Q|D^{(-1)}(e^{\circ}))\nonumber\\
& \nonumber\\
&  \quad+\int_{-\infty}^{\infty}f_{Q|E}(q|D^{(-1)}(e^{\circ}))\log_{2}\left[
P^{\prime}(q)\right]  dq.
\end{align}

\subsection{Mutual information $I(Q^{\circ};E^{\circ})$}

The mutual information about $Q{\tiny \ } $that is gained from knowledge of
$E$ is written $I(Q;E)$ and is given in Eq. (\ref{mutual I}). The
corresponding mutual information that we gain about the value of $Q^{\circ}$,
from knowledge of $E^{\circ}$, is written as $I(Q^{\circ};E^{\circ})$ and
given by
\begin{equation}
I(Q^{\circ};E^{\circ})=\int_{e_{1}^{\circ}}^{e_{2}^{\circ}}\left[  h(Q^{\circ
})-h(Q^{\circ}|e^{\circ})\right]  f_{E^{\circ}}(e^{\circ})de^{\circ
}.\label{I(Qo;Eo)}
\end{equation}
We have, using above results for $h(Q^{\circ})$ and $h(Q^{\circ}|e^{\circ})$,
that 
\begin{align}
&  h(Q^{\circ})-h(Q^{\circ}|e^{\circ})\nonumber\\
& \nonumber\\
&  =h(Q)-h(Q|D^{(-1)}(e^{\circ}))\nonumber\\
& \nonumber\\
&  +\int_{-\infty}^{\infty}\left[  f_{Q}(q)-f_{Q|E}(q|D^{(-1)}(e^{\circ
}))\right]  \log_{2}\left[  P^{\prime}(q)\right]  dq.
\end{align}
Thus 
\begin{align}
& I(Q^{\circ};E^{\circ})\nonumber\\
& \nonumber\\
& =\int_{e_{1}^{\circ}}^{e_{2}^{\circ}}\left[  h(Q^{\circ})-h(Q^{\circ
}|e^{\circ})\right]  f_{E^{\circ}}(e^{\circ})de^{\circ}\nonumber\\
& \nonumber\\
& =\int_{e_{1}^{\circ}}^{e_{2}^{\circ}}\left[  h(Q)-h(Q|D^{(-1)}(e^{\circ
}))\right]  \frac{f_{E}(D^{(-1)}(e^{\circ}))}{D^{\prime}(D^{(-1)}(e^{\circ}%
))}de^{\circ}\nonumber\\
& \nonumber\\
& \quad+\int_{-\infty}^{\infty}dq\int_{e_{1}^{\circ}}^{e_{2}^{\circ}}%
de^{\circ}\left[  f_{Q}(q)-f_{Q|E}(q|D^{(-1)}(e^{\circ}))\right]  \nonumber\\
& \nonumber\\
& \quad\times\log_{2}\left[  P^{\prime}(q)\right]  \frac{f_{E}(D^{(-1)}%
(e^{\circ}))}{D^{\prime}(D^{(-1)}(e^{\circ}))}.
\end{align}
The second integral in the above expression vanishes identically, hence
$I(Q^{\circ};E^{\circ})=\int_{-\infty}^{\infty}\left[  h(Q)-h(Q|e)\right]
f_{E}(e)de$ or
\begin{equation}
I(Q^{\circ};E^{\circ})=I(Q;E).
\end{equation}
It follows that when $Q^{\circ}$ is related to $Q$ by an increasing
transformation, and $E^{\circ}$ is related to $E$ by a generally different
increasing transformation (as given in Eqs. (\ref{P}) and (\ref{D}),
respectively), the mutual information between $Q^{\circ}$ and $E^{\circ}$ is
identical to the mutual information between $Q$ and $E$.


\section{Maximum mutual information}
\label{Maximum Appendix}


In this appendix, we vary the distribution $f_{E}(e)$ and determine a
condition that the mutual information is maximal. This condition implicitly
determines the form of $f_{E}(e)$ that maximises the mutual information. We
write the maximising form of the distribution of $E$ as $\tilde{f}_{E}(e)$.
From (1) it follows that the corresponding distribution of $Q$, when the
mutual information is maximised, is $\tilde{f}_{Q}(q)=\int f_{Q|E}
(q|e)\tilde{f}_{E}(e)de$.

The rationale of the calculations in this appendix are as follows: (i) the
mutual information is a \textit{functional} of the distribution of $E$, thus
to determine the maximum mutual information, we perform a functional change in
the distribution of $E$, such that the distribution always lies within the
space non-negative functions that have a total integral of unity; (ii) we look
for the condition that the mutual information does not change, to
\textit{linear order} in the functional change of the distribution of $E$.
This is the condition for the maximum mutual information, and leads to an
equation that determines the maximising distribution of $E$ and the channel
capacity.\bigskip

We begin, noting that in Appendix B it was shown that the mutual information
between $Q$ and $E$ is identical to the mutual information between $Q^{\circ}
$ and $E^{\circ}$. The mutual information can thus be maximised when expressed
in terms of distributions of $Q$ and $E$, or in terms of distributions of
$Q^{\circ}$ and $E^{\circ}$. We shall carry out the calculations in terms of
the distributions of $Q^{\circ}$ and $E^{\circ}$, since this is an efficient
way of obtaining all of the results we require.

We rewrite the form of the mutual information in Eq. (\ref{I(Qo;Eo)}) as
\begin{align}
& I(Q^{\circ};E^{\circ})\nonumber\\
& \nonumber\\
& =\int_{q_{1}^{\circ}}^{q_{2}^{\circ}}\int_{e_{1}^{\circ}}^{e_{2}^{\circ}%
}f_{Q^{\circ}|E^{\circ}}(q^{\circ}|e^{\circ})\left\{  \log_{2}\left[
f_{Q^{\circ}|E^{\circ}}(q^{\circ}|e^{\circ})\right]  \right.  \nonumber\\
& \nonumber\\
& \quad\left.  -\log_{2}\left[  f_{Q^{\circ}}(q^{\circ})\right]  \right\}
f_{E^{\circ}}(e^{\circ})dq^{\circ}de^{\circ}.\label{I rewrite}%
\end{align}
The above expression for $I(Q^{\circ};E^{\circ})$ depends on the form of the
distribution $f_{E^{\circ}}(e^{\circ})$ (which by Eq. (\ref{marginal Edot}) is
determined from the form of $f_{E}(e)$). We proceed by determining how
$I(Q^{\circ};E^{\circ})$ behaves under the functional change of $f_{E^{\circ}
}(e^{\circ})$ given by
\begin{equation}
f_{E^{\circ}}(e^{\circ})\rightarrow f_{E^{\circ}}(e^{\circ})+\Delta
f_{E^{\circ}}(e^{\circ})\label{change in fE}
\end{equation}

Let $\Delta f_{Q^{\circ}}(q^{\circ})$ denote the change in $f_{Q^{\circ}
}(q^{\circ})$ that is produced by the change of $\Delta f_{E^{\circ}}
(e^{\circ})$ in $f_{E^{\circ}}(e^{\circ})$. The transformed version of Eq.
(\ref{f Q}) is $f_{Q^{\circ}}(q^{\circ})=\int_{e_{1}^{\circ}}^{e_{2}^{\circ}
}f_{Q^{\circ}|E^{\circ}}(q^{\circ}|e^{\circ})f_{E^{\circ}}(e^{\circ}
)de^{\circ}$ and hence
\begin{equation}
\Delta f_{Q^{\circ}}(q^{\circ})=\int_{e_{1}^{\circ}}^{e_{2}^{\circ}
}f_{Q^{\circ}|E^{\circ}}(q^{\circ}|e^{\circ})\Delta f_{E^{\circ}}(e^{\circ
})de^{\circ}.
\end{equation}
This result indicates that $\Delta f_{Q^{\circ}}(q^{\circ})$ depends linearly
on $\Delta f_{E^{\circ}}(e^{\circ})$.

We define $\Delta I(Q^{\circ};E^{\circ})$ to be the change in the mutual
information, produced by the change of $\Delta f_{E^{\circ}}(e^{\circ})$ in
$f_{E^{\circ}}(e^{\circ})$, to precisely \textit{first order} in $\Delta
f_{E^{\circ} }(e^{\circ})$, while the change in $I(Q^{\circ};E^{\circ})$ to
second order in $\Delta f_{E^{\circ}}(e^{\circ})$ indicates that the mutual
information has a maximum (results not shown). We have
\begin{align}
& \Delta I(Q^{\circ};E^{\circ})\nonumber\\
& \nonumber\\
& =\int_{q_{1}^{\circ}}^{q_{2}^{\circ}}\int_{e_{1}^{\circ}}^{e_{2}^{\circ}%
}f_{Q^{\circ}|E^{\circ}}(q^{\circ}|e^{\circ})\left\{  \log_{2}\left[
f_{Q^{\circ}|E^{\circ}}(q^{\circ}|e^{\circ})\right]  \right.  \nonumber\\
& \nonumber\\
& \quad\left.  -\log_{2}\left[  f_{Q^{\circ}}(q^{\circ})\right]  \right\}
\Delta f_{E^{\circ}}(e^{\circ})dq^{\circ}de^{\circ}\nonumber\\
& \nonumber\\
& \quad-\int_{q_{1}^{\circ}}^{q_{2}^{\circ}}\int_{e_{1}^{\circ}}^{e_{2}%
^{\circ}}f_{Q^{\circ}|E^{\circ}}(q^{\circ}|e^{\circ})f_{E^{\circ}}(e^{\circ
})\frac{\Delta f_{Q^{\circ}}(q^{\circ})dq^{\circ}de^{\circ}}{f_{Q^{\circ}%
}(q^{\circ})\ln(2)}\nonumber\\
& \nonumber\\
& =\int_{q_{1}^{\circ}}^{q_{2}^{\circ}}\int_{e_{1}^{\circ}}^{e_{2}^{\circ}%
}f_{Q^{\circ}|E^{\circ}}(q^{\circ}|e^{\circ})\left\{  \log_{2}\left[
f_{Q^{\circ}|E^{\circ}}(q^{\circ}|e^{\circ})\right]  \right.  \nonumber\\
& \nonumber\\
& \quad\left.  -\log_{2}\left[  f_{Q^{\circ}}(q^{\circ})\right]  \right\}
\Delta f_{E^{\circ}}(e^{\circ})dq^{\circ}de^{\circ}\nonumber\\
& \nonumber\\
& \quad-\int_{q_{1}^{\circ}}^{q_{2}^{\circ}}\frac{1}{\ln(2)}\Delta
f_{Q^{\circ}}(q^{\circ})dq^{\circ}.\label{Delta I unsimplified}%
\end{align}
We note that because $\int_{e_{1}^{\circ}}^{e_{2}^{\circ}}f_{E^{\circ}
}(e^{\circ})de^{\circ}=1$ the change $\Delta f_{E^{\circ}}(e^{\circ})$ in
$f_{E^{\circ}}(e^{\circ})$ is subject to the condition
\begin{equation}
\int_{e_{1}^{\circ}}^{e_{2}^{\circ}}\Delta f_{E^{\circ}}(e^{\circ})de^{\circ
}=0.\label{int Delta}
\end{equation}
For the same reason, $\Delta f_{Q^{\circ}}(q^{\circ})$ has a vanishing
integral\footnote{We have $\Delta f_{Q^{\circ}}(q^{\circ})=\int_{e_{1}^{\circ
}}^{e_{2}^{\circ}}f_{Q^{\circ}|E^{\circ}}(q^{\circ}|e^{\circ})\Delta
f_{E^{\circ}}(e^{\circ})de^{\circ}$. Integrating this over $q^{\circ}$ and
using normalisation of $f_{Q^{\circ}|E^{\circ}}(q^{\circ}|e^{\circ})$, we
obtain $\int_{q_{1}^{\circ}}^{q_{2}^{\circ}}\Delta f_{Q^{\circ}}(q^{\circ
})dq^{\circ}=\int_{e_{1}^{\circ}}^{e_{2}^{\circ}}\Delta f_{E^{\circ}}
(e^{\circ})de^{\circ}$. Hence vanishing of $\int_{e_{1}^{\circ}}^{e_{2}
^{\circ}}\Delta f_{E^{\circ}}(e^{\circ})de^{\circ}$ produces vanishing of
$\int_{q_{1}^{\circ}}^{q_{2}^{\circ}}\Delta f_{Q^{\circ}}(q^{\circ})dq^{\circ
}$.}: $\int_{q_{1}^{\circ}}^{q_{2}^{\circ}}\Delta f_{Q^{\circ}}(q^{\circ
})dq^{\circ}=0$, which indicates that the final term in Eq.
(\ref{Delta I unsimplified}) vanishes, and $\Delta I(Q^{\circ};E^{\circ})$
reduces to
\begin{align}
\Delta I(Q^{\circ};E^{\circ})  & =\int_{q_{1}^{\circ}}^{q_{2}^{\circ}}%
\int_{e_{1}^{\circ}}^{e_{2}^{\circ}}f_{Q^{\circ}|E^{\circ}}(q^{\circ}%
|e^{\circ})\nonumber\\
& \nonumber\\
& \quad\times\left\{  \log_{2}\left[  f_{Q^{\circ}|E^{\circ}}(q^{\circ
}|e^{\circ})\right]  -\log_{2}\left[  f_{Q^{\circ}}(q^{\circ})\right]
\right\}  \nonumber\\
& \nonumber\\
& \quad\times\Delta f_{E^{\circ}}(e^{\circ})dq^{\circ}de^{\circ}%
.\label{Delta I}%
\end{align}

Let $\tilde{f}_{E^{\circ}}(e^{\circ})$ denote the form of $f_{E^{\circ}
}(e^{\circ})$ that makes $\Delta I(Q^{\circ};E^{\circ})$ vanish and maximises
the mutual information. That is, setting $f_{E^{\circ}}(e^{\circ})$ equal to
$\tilde{f}_{E^{\circ}}(e^{\circ})$ in Eq. (\ref{Delta I}) leads to the
condition
\begin{align}
&
\begin{array}
[c]{rcc}%
\int_{q_{1}^{\circ}}^{q_{2}^{\circ}}\int_{e_{1}^{\circ}}^{e_{2}^{\circ}%
}f_{Q^{\circ}|E^{\circ}}(q^{\circ}|e^{\circ})\left\{  \log_{2}\left[
f_{Q^{\circ}|E^{\circ}}(q^{\circ}|e^{\circ})\right]  \right.   &  & \\
&  & \\
\left.  -\log_{2}\left[  \tilde{f}_{Q^{\circ}}(q^{\circ})\right]  \right\}
\Delta f_{E^{\circ}}(e^{\circ})dq^{\circ}de^{\circ} & = & 0
\end{array}
\nonumber\\
& \label{general condition max}%
\end{align}
where
\begin{equation}
\tilde{f}_{Q^{\circ}}(q^{\circ})=\int_{e_{1}^{\circ}}^{e_{2}^{\circ}
}f_{Q^{\circ}|E^{\circ}}(q^{\circ}|e^{\circ})\tilde{f}_{E^{\circ}}(e^{\circ
})de^{\circ}\label{f tilde Q0}
\end{equation}
is the marginal distribution of $Q^{\circ}$ when $E^{\circ}$ has the
maximising distribution, $\tilde{f}_{E^{\circ}}(e^{\circ})$. Generally, we
indicate distributions that depend on the mutual-information maximising
distribution of $E$ by a tilde\footnote{Thus in Eq. (\ref{f tilde Q0}), the
distribution of $Q^{\circ}$, namely $\tilde{f}_{Q^{\circ}}(q^{\circ})$, has
been decorated with a tilde because it depends on the distribution $\tilde
{f}_{E^{\circ}}(e^{\circ})$, which, in turn, depends on the distribution
$\tilde{f}_{E}(e)$, which maximises the mutual information.}.

Apart from $\Delta f_{E^{\circ}}(e^{\circ})$ satisfying Eq. (\ref{int Delta}),
it can be chosen in an arbitrary way\footnote{More precisely, apart from
$\Delta f_{E}(e)$ satisfying Eq. (\ref{int Delta}), it must not cause the
distribution of $E$ to become negative for any $e$, but is otherwise
arbitrary.}. The most general way for Eq. (\ref{general condition max}) to
hold is for the coefficient of $\Delta f_{E^{\circ}}(e^{\circ})$ in Eq.
(\ref{general condition max}) to equal a constant that we shall write as $C$.
This general condition follows since then the right hand side of Eq.
(\ref{Delta I}) takes the form $\int_{e_{1}^{\circ}}^{e_{2}^{\circ}}C\Delta
f_{E^{\circ}}(e^{\circ})de^{\circ}$ which vanishes identically because of Eq.
(\ref{int Delta}), resulting in $\Delta I(Q^{\circ};E^{\circ})=0$. Thus the
condition for the mutual information to be maximised at $f_{E^{\circ}
}(e^{\circ})=\tilde{f}_{E^{\circ}}(e^{\circ})$ is
\begin{align}
&
\begin{array}
[c]{lcr}%
\int_{q_{1}^{\circ}}^{q_{2}^{\circ}}f_{Q^{\circ}|E^{\circ}}(q^{\circ}%
|e^{\circ}) &  & \\
&  & \\
\times\left\{  \log_{2}\left[  f_{Q^{\circ}|E^{\circ}}(q^{\circ}|e^{\circ
})\right]  -\log_{2}\left[  \tilde{f}_{Q^{\circ}}(q^{\circ})\right]  \right\}
dq^{\circ} & = & C.
\end{array}
\nonumber\\
& \label{condition max}%
\end{align}

Equation (\ref{condition max}) is an equation that implicitly determines: (i)
the distribution of $E^{\circ}$ that maximises the mutual information, namely
$\tilde{f}_{E^{\circ}}(e^{\circ})$, and (ii) the constant $C$. In Appendix
\ref{Slow Change Appendix} we give an approximate analysis that illustrates
how both $\tilde{f}_{E^{\circ}}(e^{\circ})$ and $C$ are determined from Eq.
(\ref{condition max}).

When the distribution $f_{E^{\circ}}(e^{\circ})$ is set equal to $\tilde
{f}_{E^{\circ}}(e^{\circ})$ within $I(Q^{\circ},E^{\circ})$, the result is the
maximum mutual information, which we write as $\widetilde{I}(Q^{\circ
},E^{\circ})$. Using Eq. (\ref{I rewrite}), we can write $\widetilde{I}
(Q^{\circ},E^{\circ})$ as 
$\int_{q_{1}^{\circ}}^{q_{2}^{\circ}}\int_{e_{1}^{\circ}}^{e_{2}^{\circ}%
}f_{Q^{\circ}|E^{\circ}}(q^{\circ}|e^{\circ})$\newline
$\times\left\{  \log_{2}\left[  f_{Q^{\circ}|E^{\circ}}(q^{\circ}|e^{\circ
})\right]  -\log_{2}\left[  \tilde{f}_{Q^{\circ}}(q^{\circ})\right]  \right\}
\tilde{f}_{E^{\circ}}(e^{\circ})dq^{\circ}de^{\circ}$
and using Eq.
(\ref{condition max}) within this expression yields $\int_{e_{1}^{\circ}
}^{e_{2}^{\circ}}C~\tilde{f}_{E^{\circ}}(e^{\circ})de^{\circ}=C$. Hence we
have
\begin{equation}
\widetilde{I}(Q^{\circ},E^{\circ})=\text{maximum mutual information}=C.
\end{equation}
Thus, the constant $C$ in Eq. (\ref{condition max}) represents the maximum
value of the mutual information, i.e., the \textit{channel capacity}
(\cite{Shannon}, \cite{CoverThomas}).

Note that by simply taking the transformations $P(x)$ and $D(x)$ of Eqs.
(\ref{P}) and (\ref{D}), respectively, to be the identity transformation:
$P(x)=x$ and $D(x)=x$, leads to versions of Eqs. (\ref{f tilde Q0}) and
(\ref{condition max}) that apply to the original variables. That is
\begin{equation}
\tilde{f}_{Q}(q)=\int_{-\infty}^{\infty}f_{Q|E}(q|e)\tilde{f}_{E}
(e)de\label{f tilde Q}
\end{equation}
and
\begin{equation}
\int_{-\infty}^{\infty}f_{Q|E}(q|e)\left\{  \log_{2}\left[  f_{Q|E}
(q|e)\right]  -\log_{2}\left[  \tilde{f}_{Q}(q)\right]  \right\}
dq=C\label{condition max Q E}
\end{equation}
where $\tilde{f}_{E}(e)$ is the distribution of $E$ that maximises the mutual
information, $I(Q,E)$, and is related to $\tilde{f}_{E^{\circ}}(e^{\circ})$ by
Eq. (\ref{marginal Edot}).


\section{Special transformed variables}
\label{Special Appendix}

In this appendix, we introduce a special transformed version of the random
variable $Q$ whose entropy is \textit{homogenized} in the sense it is
independent of any value that $E$ is conditioned upon. We additionally
introduce two special transformed versions of $E$.

Let us first consider the form of Eq. (\ref{condition max}) that applies when
the increasing transformation $P(x)$, that appears in Eq. (\ref{P}), has the
\textit{special form}
\begin{equation}
P(x)=\tilde{F}_{Q}(x)\label{special P}
\end{equation}
where
\begin{equation}
\tilde{F}_{Q}(x)=\int_{-\infty}^{x}\tilde{f}_{Q}(q)dq=\int_{-\infty}^{x}
dq\int_{-\infty}^{\infty}de\,f_{Q|E}(q|e)\tilde{f}_{E}(e).\label{F tilde Q}
\end{equation}
is the cumulative distribution function of $Q$ when $f_{E}(e)$ is the mutual
information maximising distribution $\tilde{f}_{E}(e)$. Since we are
considering a special transformation, we shall give the transformed variable a
special name and call it $Q^{\ast}$ (rather than $Q^{\circ}$). Thus, we
define
\begin{equation}
Q^{\ast}=\tilde{F}_{Q}(Q)
\end{equation}
(cf. Eq. (\ref{P})). Taking into account that $Q^{\ast}$ can only take values
in the range $0$ to $1$ (because $\tilde{F}_{Q}(x)$ is a cumulative
distribution), the integral in Eq. (\ref{condition max}) covers the range $0$
to $1$. Additionally, the special choice of $P(x)$ in Eq. (\ref{special P})
causes $\tilde{f}_{Q^{\ast}}(q^{\ast})$ to take the value of unity over the
range $0$ to $1$ of the integral\footnote{The distribution $\tilde{f}
_{Q^{\ast}}(q^{\ast})$ follows from Eq. (\ref{marginal Qdot}) with: (i)
$f_{Q}(x)$ set equal to $\tilde{f}_{Q}(x)$ (given in Eq. (\ref{f tilde Q})),
(ii) $P(x)$ set equal to $\tilde{F}_{Q}(x)$ (given in Eq. (\ref{F tilde Q})).
Then for $0<q^{\ast}<1$ we have $\tilde{f}_{Q^{\ast}}(q^{\ast})=\dfrac
{\tilde{f}_{Q}(\tilde{F}_{Q}^{(-1)}(q^{\ast}))}{\tilde{F}_{Q}{}^{\prime
}(\tilde{F}_{Q}^{(-1)}(q^{\ast}))}$ and since $\tilde{F}_{Q}^{\prime
}(x)=\tilde{f}_{Q}(x)$ we have $\tilde{f}_{Q^{\ast}}(q^{\ast})=1$.}. 
With this form of $\tilde{f}_{Q^{\ast}}(q^{\ast})$, Eq. (\ref{condition max}) becomes
\begin{equation}
\int_{0}^{1}f_{Q^{\ast}|E^{\circ}}(q^{\ast}|e^{\circ})\log_{2}\left[
f_{Q^{\ast}|E^{\circ}}(q^{\ast}|e^{\circ})\right]  dq^{\ast}
=C.\label{special condition}
\end{equation}
Equation (\ref{special condition}) can be written in the compact form
\begin{equation}
h(Q^{\ast}|e^{\circ})=-C.\label{h(Q*|e0)}
\end{equation}
This result signals homogenization of the entropy, where the entropy of
$Q^{\ast}$ is independent of the value that $E^{\circ}$ (and hence $E$) is
conditioned upon.

We note that the transformation from $E$ to $E^{\circ}$, namely $D(x)$, is
arbitrary. The special choice $D(x)=\tilde{F}_{E}(x)$ where
\begin{equation}
\tilde{F}_{E}(x)=\int_{-\infty}^{x}\tilde{f}_{E}(e)de\label{F tilde E}
\end{equation}
leads to a transformed variable we call $E^{\ast}$ which is defined by
\begin{equation}
E^{\ast}=\tilde{F}_{E}(E).
\end{equation}
The variable $E^{\ast}$, like $Q^{\ast}$, has a uniform distribution:
$\tilde{f}_{E^{\ast}}(e^{\ast})=1$ for $0<e^{\ast}<1$ (and is zero elsewhere)
and Eq. (\ref{h(Q*|e0)}) takes the form
\begin{equation}
h(Q^{\ast}|e^{\ast})=-C.\label{-h* = C}
\end{equation}
Another special choice for $D(x)$ is $D(x)=x$ and leads to $E^{\circ}=E$ and
we write the corresponding version of Eq. (\ref{h(Q*|e0)}) as $h(Q^{\ast
}|e)=-C$.


\section{Slow change regime}
\label{Slow Change Appendix}


In this appendix, approximate results are derived, based on the assumption
that $f_{Q|E}(q|e)$ has, as a function of $q$, a width that is very small for
all $e$,

We shall now work under the explicit assumption both $q$ and $e$ lie in a
finite range of values, given by
\begin{equation}
q_{\min}<q<q_{\max}\text{ and }e_{\min}<e<_{\max}.\label{finite range}
\end{equation}

\subsection{Approximate analysis for the slow change regime}

The approximate analysis we shall present is based on the key assumption that
the distribution $f_{Q|E}(q|e)$ has, as a function of $q$, a width that is
very small for all $e$. To specify this more precisely, we note that given Eq.
(\ref{finite range}) we have
\begin{equation}
\int_{q_{\min}}^{q_{\max}}f_{Q|E}(q|e)dq=1\label{norm fqe}
\end{equation}
which indicates that $f_{Q|E}(q|e)$, as a function of $q$, is a normalised
probability density. We define the mean and variance of $f_{Q|E}(q|e)$,
written $m(e)$ and $\sigma^{2}(e)$, respectively, as
\begin{align}
m(e) &  =\int_{q_{\min}}^{q_{\max}}qf_{Q|E}(q|e)dq\label{mean fqe}\\
& \nonumber\\
\sigma^{2}(e) &  =\int_{q_{\min}}^{q_{\max}}\left[  q-m(e)\right]  ^{2}
f_{Q|E}(q|e)dq\label{var fqe}
\end{align}
and take $\sigma(e)$ to be positive.

Since $q_{\min}<q<q_{\max}$ we must have $m(e)$ also lying somewhere between
$q_{\min}$ and $q_{\max}$. We assume that $Q$ and $E$ are positively
correlated by taking $m(e)$ to be an increasing function of $e$, i.e.,
$m^{\prime}(e)>0$ where $m^{\prime}(e)=dm(e)/de$. We thus have
\begin{equation}
q_{\min}<m(0)<m(1)<q_{\max}.\label{mrange}
\end{equation}

Let us write $f_{Q|E}(q|e)$ in terms of a new function $\phi\left(
x|e\right)  $ defined by
\begin{equation}
f_{Q|E}(q|e)=\frac{1}{\sigma(e)}\phi\left(  \left.  \frac{q-m(e)}{\sigma
(e)}\right\vert e\right)  .\label{fqe = phi}
\end{equation}
The properties of $f_{Q|E}(q|e)$ in Eq. (\ref{norm fqe}) - (\ref{var fqe} )
are fully reproduced when $\phi\left(  x|e\right)  $ has the properties
\begin{align}
\int_{x_{1}}^{x_{2}}\phi\left(  x|e\right)  dx &  =1\label{norm phi}\\
& \nonumber\\
\int_{x_{1}}^{x_{2}}x\phi\left(  x|e\right)  dx &  =0\label{mean phi}\\
& \nonumber\\
\int_{x_{1}}^{x_{2}}x^{2}\phi\left(  x|e\right)  dx &  =1\label{var phi}
\end{align}
where
\begin{equation}
x_{1}=\frac{q_{\min}-m(e)}{\sigma(e)}\text{ and }x_{2}=\frac{q_{\max}
-m(e)}{\sigma(e)}.\label{xm xp}
\end{equation}

We write the maximum value of $\sigma(e)$, over all $e$, as $\sigma_{\max}$,
hence
\begin{equation}
\sigma(e)\leq\sigma_{\max}.
\end{equation}
We make approximations for the regime where $\sigma_{\max}$ is small. More
explicitly, when $e$ changes by an amount $\sigma_{\max}$, i.e. $e\rightarrow
e+\sigma_{\max}$ we have $m(e+\sigma_{\max})\simeq m(e)\times\left(
1+\sigma_{\max}\frac{m^{\prime}(e)}{m(e)}\right)  $ and the quantity
$\sigma_{\max}\frac{m^{\prime}(e)}{m(e)}\equiv\sigma_{\max}\frac{d\ln
(m(e))}{de}$ is a measure of the \textit{fractional} change in $m(e)$. The
slow change regime corresponds to small fractional changes in $m(e)$,
$\phi\left(  x|e\right)  $, and $\sigma(e)$ occurring when $e$ changes by
$\sigma_{\max}$, i.e., $\sigma_{\max}\frac{d\ln(m(e))}{de}\ll1$, $\sigma
_{\max}\frac{\partial\ln(\phi\left(  x|e\right)  )}{\partial e}\ll1$, and
$\sigma_{\max}\frac{d\ln(\sigma(e))}{de}\ll1$.

\subsection{Approximation of the distribution $\tilde{f}_{E}(e)$ in the slow
change regime}

We shall now present results when the \textit{mutual information is
maximised}. That is, where $E$ is governed by the information-maximising
distribution $\tilde{f}_{E}(e)$, and as a consequence, the marginal
distribution of $Q$ is $\tilde{f}_{Q}(q)$.

To find an approximation for $\tilde{f}_{E}(e)$ we start with the condition
that mutual information is maximised, Eq. (\ref{condition max Q E}), which
applies for all allowed values of $e$. We can write this equation, with no
approximation, as
\begin{equation}
-\int_{q_{\min}}^{q_{\max}}f_{Q|E}(q|e)\log_{2}\left[  \tilde{f}
_{Q}(q)\right]  dq=C+h(Q|e)\label{int fqe log fq}
\end{equation}
where $h(Q|e)$ is given by Eq. (\ref{h(Q|e)}).

In terms of $\phi\left(  x|e\right)  $ we can write Eq. (\ref{int fqe log fq})
as
\begin{equation}%
\begin{array}
[c]{l}%
-\int_{q_{\min}}^{q_{\max}}\frac{1}{\sigma(e)}\phi\left(  \left.
\frac{q-m(e)}{\sigma(e)}\right\vert e\right)  \log_{2}\left[  \tilde{f}%
_{Q}(q)\right]  dq\\
\\
\quad=C+h(Q|e)
\end{array}
\end{equation}
and using the integration variable $x=\frac{q-m(e)}{\sigma(e)}$ yields
\begin{equation}
-\int_{x_{1}}^{x_{2}}\phi\left(  x|e\right)  \log_{2}\left[  \tilde{f}
_{Q}(m(e)+\sigma(e)x)\right]  dx=C+h(Q|e).
\end{equation}
In this expression, $x$ is effectively restricted to a range of $O(1)$,
because of Eqs. (\ref{norm phi}) and (\ref{var phi}). We assume that
$\tilde{f}_{Q}(q)$ changes very little over an interval of $\sigma_{\max}$,
so\footnote{The result we obtain later, for $\tilde{f}_{Q}(q)$ is consistent
with this assumption.} the leading approximation of the above equation, for
small $\sigma_{\max}$, follows from neglecting the $\sigma(e)x$ term on the
left hand side. This leads to $-\int_{x_{1}}^{x_{2}}\phi\left(  x|e\right)
\log_{2}\left[  \tilde{f}_{Q}(m(e))\right]  dx\simeq C+h(Q|e)$. Using Eq.
(\ref{norm phi}), this equation reduces to $-\log_{2}\left[  \tilde{f}
_{Q}(m(e))\right]  \simeq C+h(Q|e)$ or
\begin{equation}
\tilde{f}_{Q}(m(e))\simeq\frac{1}{2^{C+h(Q|e)}}.\label{fQ(m)=1/2^}
\end{equation}

We can obtain another expression for $\tilde{f}_{Q}(m(e))$ that holds under
similar conditions. We have, by definition, that $\tilde{f}_{Q}(q)=\int
_{e_{\min}}^{e_{\max}}f_{Q|E}(q|r)\tilde{f}_{E}(r)dr$ which can be written in
terms of $\phi\left(  x|r\right)  $ as $\tilde{f}_{Q}(q)=\int_{e_{\min}
}^{e_{\max}}\frac{1}{\sigma(r)}\phi\left(  \left.  \frac{q-m(r)}{\sigma
(r)}\right\vert r\right)  \tilde{f}_{E}(r)dr$. Setting $q=m(e)$ in this
expression gives
\begin{equation}
\tilde{f}_{Q}(m(e))=\int_{e_{\min}}^{e_{\max}}\frac{1}{\sigma(r)}\phi\left(
\left.  \frac{m(e)-m(r)}{\sigma(r)}\right\vert r\right)  \tilde{f}_{E}(r)dr.
\end{equation}
Given the properties of $\phi\left(  x|r\right)  $, the above integral is
dominated by the range of $r$ given by $\left\vert \frac{m(e)-m(r)}{\sigma
(r)}\right\vert \lesssim1$ and working under the assumption that $\tilde
{f}_{E}(r)$ varies slowly with $r$ we have
\begin{align}
\tilde{f}_{Q}(m(e)) &  \simeq\frac{\tilde{f}_{E}(e)}{\sigma(e)}\int_{e_{\min}
}^{e_{\max}}\phi\left(  \left.  \frac{m(e)-m(r)}{\sigma(e)}\right\vert
e\right)  dr\nonumber\\
& \nonumber\\
&  \simeq\frac{\tilde{f}_{E}(e)}{\sigma(e)}\int_{e_{\min}}^{e_{\max}}
\phi\left(  \left.  \frac{-(r-e)m^{\prime}(e)}{\sigma(e)}\right\vert e\right)
dr\nonumber\\
& \nonumber\\
&  \simeq\frac{\tilde{f}_{E}(e)}{m^{\prime}(e)}.\label{fQ(m) = fE/m'}
\end{align}

Comparing Eqs. (\ref{fQ(m)=1/2^}) and (\ref{fQ(m) = fE/m'}) yields
\begin{equation}
\tilde{f}_{E}(e)\simeq\frac{m^{\prime}(e)}{2^{C+h(Q|e)}}. \label{unnormalised}
\end{equation}
This is the approximate distribution of $E$ that maximises the mutual
information. Additionally, from Eq. (\ref{fQ(m) = fE/m'}), the approximate
distribution of $Q$ that applies when the mutual information is maximised is
\begin{align}
\tilde{f}_{Q}(q)  & \simeq\frac{\tilde{f}_{E}(m^{(-1)}(q))}{m^{\prime
}(m^{(-1)}(q))}\nonumber\\
& \nonumber\\
& =\frac{m^{\prime}(m^{(-1)}(q))}{m^{\prime}(m^{(-1)}(q))2^{C+h(Q|m^{(-1)}%
(q))}}\nonumber\\
& \nonumber\\
& =\frac{1}{2^{C+h(Q|m^{(-1)}(q))}}.
\end{align}

Since $\tilde{f}_{E}(e)$ is normalised to unity, we can infer the channel
capacity from $1=\int_{e_{\min}}^{e_{\max}}\tilde{f}_{E}(e)de\simeq
\int_{e_{\min}}^{e_{\max}}\frac{m^{\prime}(e)}{2^{C+h(Q|e)}}de$. This yields
\begin{equation}
C\simeq\log_{2}\left(  \int_{e_{\min}}^{e_{\max}}\frac{m^{\prime}
(e)}{2^{h(Q|e)}}de\right)  .
\end{equation}
We can also write $\tilde{f}_{E}(e)$ as
\begin{equation}
\tilde{f}_{E}(e)\simeq\frac{1}{N}\times\frac{m^{\prime}(e)}{2^{h(Q|e)}}
\end{equation}
where
\begin{equation}
N=\int_{e_{\min}}^{e_{\max}}\frac{m^{\prime}(x)}{2^{h(Q|x)}}dx.
\end{equation}

We note that the above results apply when $q$ and $e$ have \textit{finite
ranges}. A finite range of $q$ and $e$ was adopted since for variables with an
infinite range, the form of $\tilde{f}_{E}(e)$ in Eq. (\ref{unnormalised}) is
not guaranteed to be normalisable.

\subsection{Particular results for the slow change regime}

We shall now determine some approximate results in the slow change regime when
$E$ has the mutual-information-maximising distribution $\tilde{f}_{E}(e) $.
The transformations adopted for $Q$ and $E$ are
\begin{equation}
Q^{\ast}=\tilde{F}_{Q}(Q)\text{ and }E^{\ast}=\tilde{F}_{E} (E)
\label{special transforms}
\end{equation}
where $\tilde{F}_{Q}(x)$ and $\tilde{F}_{E}(e)$ are given in Eqs.
(\ref{f tilde Q}) and (\ref{F tilde E}), respectively.

The joint distribution $f_{Q^{\ast},E^{\ast}}(q^{\ast},e^{\ast})$ is given by
\begin{align}
f_{Q^{\ast},E^{\ast}}(q^{\ast},e^{\ast})  & =\int_{q_{\min}}^{q_{\max}}%
dq\int_{e_{\min}}^{e_{\max}}de\delta\left(  q^{\ast}-\tilde{F}_{Q}(q)\right)
\nonumber\\
& \nonumber\\
& \quad\times\delta\left(  e^{\ast}-\tilde{F}_{E}(q)\right)  f_{Q|E}%
(q|e)\tilde{f}_{E}(e).
\end{align}
For $0<q^{\ast}<1$ and $0<e^{\ast}<1$ we have $f_{Q^{\ast},E^{\ast}}(q^{\ast
},e^{\ast})$ non-zero and given by
\begin{equation}
f_{Q^{\ast},E^{\ast}}(q^{\ast},e^{\ast})=\frac{f_{Q|E}(\tilde{F}_{Q}
^{(-1)}(q^{\ast})|\tilde{F}_{E}^{(-1)}(e^{\ast}))}{\tilde{f}_{Q}\left(
\tilde{F}_{Q}^{(-1)}(q^{\ast})\right)  }.\label{fQ*E*}
\end{equation}
From the results in Appendix \ref{Special Appendix}, we know that $\tilde
{f}_{Q^{\ast}}(q^{\ast})$ and $\tilde{f}_{E^{\ast}}(e^{\ast})$ are uniform
distributions on $0$ to $1$ (and zero elsewhere). This has the consequence
that for $q^{\ast}$ and $e^{\ast}$ both in the range $0$ to $1$ that
\begin{equation}
\tilde{f}_{Q^{\ast},E^{\ast}}(q^{\ast},e^{\ast})=f_{Q^{\ast}|E^{\ast}}
(q^{\ast}|e^{\ast})=\tilde{f}_{E^{\ast}|Q^{\ast}}(e^{\ast}|q^{\ast
}).\label{same}
\end{equation}

\subsection{Conditional mean values of $Q^{\ast}$ and $E^{\ast}$ in the slow
change regime}

The mean value of $Q^{\ast}$, conditional on $E^{\ast}=e^{\ast}$, is $\int
_{0}^{1}q^{\ast}\tilde{f}_{Q^{\ast}|E^{\ast}}(q^{\ast}|e^{\ast})dq^{\ast}$.
Using Eqs. (\ref{fQ*E*}) and (\ref{same}), and changing variable from
$q^{\ast}$ to $q$ via $q^{\ast}=\tilde{F}_{Q}(q)$, we obtain
\begin{align}
& \int_{0}^{1}q^{\ast}\tilde{f}_{Q^{\ast}|E^{\ast}}(q^{\ast}|e^{\ast}%
)dq^{\ast}\nonumber\\
& \nonumber\\
& =\int_{q_{\min}}^{q_{\max}}\tilde{F}_{Q}(q)f_{Q|E}(q|\tilde{F}_{E}%
^{(-1)}(e^{\ast}))dq.\label{EQ*}%
\end{align}
Similarly, the mean value of $E^{\ast}$, conditional on $Q^{\ast}=q^{\ast}$,
is $\int_{0}^{1}e^{\ast}\tilde{f}_{E^{\ast}|Q^{\ast}}(e^{\ast}|q^{\ast
})de^{\ast}$ and using Eqs. (\ref{fQ*E*}) and (\ref{same}) we find
\begin{align}
& \int_{0}^{1}e^{\ast}\tilde{f}_{E^{\ast}|Q^{\ast}}(e^{\ast}|q^{\ast}%
)de^{\ast}\nonumber\\
& \nonumber\\
& =\int_{e_{\min}}^{e_{\max}}\tilde{F}_{E}(e)\frac{f_{Q|E}(\tilde{F}%
_{Q}^{(-1)}(q^{\ast})|e)\tilde{f}_{E}\left(  e\right)  }{\tilde{f}_{Q}\left(
\tilde{F}_{Q}^{(-1)}(q^{\ast})\right)  }de.\label{EE*}%
\end{align}
We now approximate both of these results by using the assumption that
$f_{Q|E}(q|e)$ is very sharply peaked around $q=m(e)$ or equivalently around
$e=m^{(-1)}(q)$. In $\tilde{F}_{Q}(q)$ in Eq. (\ref{EQ*}), we neglect
deviations of $q$ around the mean value of $f_{Q|E}(q|\tilde{F}_{E}
^{(-1)}(e^{\ast}))$, namely $m\left(  \tilde{F}_{E}^{(-1)}(e^{\ast})\right)
$. This leads to
\begin{equation}
\int_{0}^{1}q^{\ast}\tilde{f}_{Q^{\ast}|E^{\ast}}(q^{\ast}|e^{\ast})dq^{\ast
}\simeq\tilde{F}_{Q}\left(  m\left(  \tilde{F}_{E}^{(-1)}(e^{\ast})\right)
\right).\label{EQ* app1}
\end{equation}
Similarly, in Eq. (\ref{EE*}) we obtain
\begin{equation}
\int_{0}^{1}e^{\ast}\tilde{f}_{E^{\ast}|Q^{\ast}}(e^{\ast}|q^{\ast})de^{\ast
}\simeq\tilde{F}_{E}\left(  m^{(-1)}\left(  \tilde{F}_{Q}^{(-1)}(q^{\ast
})\right)  \right).\label{EE* app1}
\end{equation}
Lastly, we write Eq. (\ref{fQ(m) = fE/m'}) in the form $\tilde{f}
_{Q}(m(e))m^{\prime}(e)\simeq\tilde{f}_{E}(e)$ and integrate to obtain
$\tilde{F}_{Q}(m(e))\simeq\tilde{F}_{E}(e)$ and using this result in Eqs.
(\ref{EQ* app1}) and (\ref{EE* app1}) yields the results
\begin{equation}
\int_{0}^{1}q^{\ast}\tilde{f}_{Q^{\ast}|E^{\ast}}(q^{\ast}|e^{\ast})dq^{\ast
}\simeq e^{\ast}\label{EQ*|E*}
\end{equation}
and
\begin{equation}
\int_{0}^{1}e^{\ast}\tilde{f}_{E^{\ast}|Q^{\ast}}(e^{\ast}|q^{\ast})de^{\ast
}\simeq q^{\ast}\label{EE*|Q*}
\end{equation}
respectively.

Equation (\ref{EQ*|E*}) tells us that with corrections of order $\sigma_{\max
}$, the expected value of $Q^{\ast}$, conditional on the value of $E^{\ast}$,
approximately equals the value that $E^{\ast}$ is conditioned upon. Equation
(\ref{EE*|Q*}) tells us that an equivalent result is obtained, when the roles
of $Q^{\ast}$ and $E^{\ast}$ are interchanged.

\subsection{Approximate property of $h(E^{\ast}|q^{\ast})$}

We have already established that $h(Q^{\ast}|e^{\ast})$ is independent of
$e^{\ast}$ (see Eq. (\ref{-h* = C})). Here we show that when the distribution
of $E$ maximises the mutual information, and the slow-change regime applies,
that $h(E^{\ast}|q^{\ast})$ is approximately independent of $q^{\ast}$.

To proceed we write Eq. (\ref{condition max Q E}) in terms of $\phi\left(
\left.  x\right\vert e\right)  $ defined in Eq. (\ref{fqe = phi}), and again
approximate $\tilde{f}_{Q}(m(e)+\sigma(e)x)$ by $\tilde{f}_{Q}(m(e))$. The
result is
\begin{equation}
\int\phi\left(  \left.  x\right\vert e\right)  \log_{2}\left[  \phi\left(
\left.  x\right\vert e\right)  \right]  dx-\log_{2}\left[  \sigma(e)\tilde
{f}_{Q}(m(e))\right]  \simeq C \label{phi eq 1}
\end{equation}
and to avoid additional notation, we have omitted integration limits, knowing
they are adequate to capture the full weight of $\phi\left(  \left.
x\right\vert e\right)  $ (see Eqs. (\ref{norm phi}) - (\ref{var phi})).
Equation (\ref{phi eq 1}) tells us that with corrections of order
$\sigma_{\max}$, the left hand side is independent of the value of $e$.

Now consider $h(E^{\ast}|q^{\ast})$. From Eq. (\ref{same}) we have that for
$q^{\ast}$ and $e^{\ast}$ both in the range $0$ to $1$ that $\tilde
{f}_{Q^{\ast},E^{\ast}}(q^{\ast},e^{\ast})=\tilde{f}_{E^{\ast}|Q^{\ast}
}(e^{\ast}|q^{\ast})$. We can thus write $h(E^{\ast}|q^{\ast})=-\int_{0}
^{1}f_{Q^{\ast},E^{\ast}}(q^{\ast},e^{\ast})\log_{2}\left[  f_{Q^{\ast
},E^{\ast}}(q^{\ast},e^{\ast})\right]  de^{\ast}$ and using Eq. (\ref{fQ*E*})
yields
\begin{align}
h(E^{\ast}|q^{\ast}) &  =-\int_{0}^{1}\tfrac{f_{Q|E}(\tilde{F}_{Q}%
^{(-1)}(q^{\ast})|\tilde{F}_{E}^{(-1)}(e^{\ast}))}{\tilde{f}_{Q}\left(
\tilde{F}_{Q}^{(-1)}(q^{\ast})\right)  }\nonumber\\
& \nonumber\\
&  \quad\times\log_{2}\left[  \tfrac{f_{Q|E}(\tilde{F}_{Q}^{(-1)}(q^{\ast
})|\tilde{F}_{E}^{(-1)}(e^{\ast}))}{\tilde{f}_{Q}\left(  \tilde{F}_{Q}%
^{(-1)}(q^{\ast})\right)  }\right]  de^{\ast}\nonumber\\
& \nonumber\\
&  =-\int_{e_{\min}}^{e_{\max}}\tfrac{f_{Q|E}(\theta|e)}{\tilde{f}_{Q}\left(
\theta\right)  }\log_{2}\left[  \tfrac{f_{Q|E}(\theta|e)}{\tilde{f}_{Q}\left(
\theta\right)  }\right]  \tilde{f}_{E}\left(  e\right)  de
\end{align}
where
\begin{equation}
\theta=\tilde{F}_{Q}^{(-1)}(q^{\ast}).
\end{equation}
In terms of $\phi\left(  \left.  x\right\vert e\right)  $ defined in Eq.
(\ref{fqe = phi}) we have
\begin{align}
h(E^{\ast}|q^{\ast}) &  =-\int_{e_{\min}}^{e_{\max}}\tfrac{1}{\tilde{f}%
_{Q}\left(  \theta\right)  }\tfrac{1}{\sigma(e)}\phi\left(  \left.
\tfrac{\theta-m(e)}{\sigma(e)}\right\vert e\right)  \nonumber\\
& \nonumber\\
&  \quad\times\log_{2}\left[  \tfrac{\frac{1}{\sigma(e)}\phi\left(  \left.
\frac{\theta-m(e)}{\sigma(e)}\right\vert e\right)  }{\tilde{f}_{Q}\left(
\theta\right)  }\right]  \tilde{f}_{E}\left(  e\right)  de\nonumber\\
& \nonumber\\
&  \simeq-\int_{e_{\min}}^{e_{\max}}\tfrac{1}{\tilde{f}_{Q}\left(
\theta\right)  }\tfrac{1}{\sigma(\alpha)}\phi\left(  \left.  \tfrac
{\theta-m(e)}{\sigma(\alpha)}\right\vert \alpha\right)  \nonumber\\
& \nonumber\\
&  \quad\times\log_{2}\left[  \tfrac{\frac{1}{\sigma(e)}\phi\left(  \left.
\tfrac{\theta-m(e)}{\sigma(\alpha)}\right\vert \alpha\right)  }{\tilde{f}%
_{Q}\left(  \theta\right)  }\right]  \tilde{f}_{E}\left(  \alpha\right)  de
\end{align}
where slowness of the $e$ dependence of various quantities has allowed us to
replace $e$ by
\begin{equation}
\alpha=m^{(-1)}(\theta).
\end{equation}
Using Eq. (\ref{fQ(m) = fE/m'}) to replace $\tilde{f}_{Q}\left(
\theta\right)  \equiv\tilde{f}_{Q}\left(  m(\alpha)\right)  $ with $\tilde
{f}_{E}(\alpha)/m^{\prime}(\alpha)$ and changing to the integration variable
$x=\tfrac{\theta-m(e)}{\sigma(\alpha)}$ yields
\begin{align}
h(E^{\ast}|q^{\ast})  & \simeq-\int\phi\left(  \left.  x\right\vert
\alpha\right)  \log_{2}\left[  \phi\left(  \left.  x\right\vert \alpha\right)
\right]  dx\nonumber\\
& \nonumber\\
& \quad+\log_{2}\left[  \sigma(\alpha)\tilde{f}_{Q}\left(  m(\alpha\right)
)\right].
\end{align}
A comparison of this result with the left hand side of Eq. (\ref{phi eq 1})
indicates that $h(E^{\ast}|q^{\ast})$ is approximately independent of $\alpha$
which is equivalent to it being independent of $q^{\ast}$.


\section{Miscellaneous results associated with\newline Figure 3}
\label{Figure 3 Appendix}


In this appendix, we give some additional results associated with Figure 3.

\subsection{Figure 3a and 3b}

For Fig. 3, we assumed that only two forms of the conditional distribution
$f_{Q|E}(q|e)$ occurred. The first form of $f_{Q|E}(q|e)$ is associated with
$e<3/5$. It is a uniform distribution that is non-zero for $q$ ranging from
$1/4$ to $3/4$, and has mean $1/2$ (see Fig. 3a). The second form of
$f_{Q|E}(q|e)$ is associated with $e\geq3/5$. It is also a uniform
distribution that is non-zero for $q$ ranging from $0$ to $1$, and also has
mean $1/2$ (see Fig. 3b). However the variance of this form of the
distribution is four times that of the first form. Here, we show that no
\textit{increasing transformation} (as defined in this work) exists that can
act on $Q$ and homogenize the variance in the sense that, after
transformation, the resulting two forms of the conditional distribution have
the same variance.

To begin, we note that Fig. 3a contains the distribution of the random
variable $Q_{1}=\frac{1}{4}+\frac{1}{2}U$ where $U$ is a random variable that
is uniformly distributed from $0$ to $1$. By contrast, Fig. 3b contains the
distribution of the random variable $Q_{2}=U$. We shall interpolate between
$Q_{1}$ and $Q_{2}$ using
\begin{equation}
Q_{b}=\frac{2-b}{4}+\frac{b}{2}U\text{ }\quad\text{with }1\leq b\leq2
\end{equation}
such that $Q_{b}$ coincides with $Q_{1}$ or $Q_{2}$, when $b=1$ or $2$, respectively.

We write the extreme values that $Q_{b}$ can take as
\begin{equation}
q_{\pm}\equiv q_{\pm}(b)=\frac{1}{2}\pm\frac{b}{4}.
\end{equation}
The random variable $Q_{b}$ is uniformly distributed from $q_{-}(b)$ to
$q_{+}(b)$, and has a height of $1/\left[  q_{+}(b)-q_{-}(b)\right]  =2/b$.

The distribution of $Q_{b}$, when evaluated at $q$, is
\begin{equation}
f_{Q_{b}}(q)=\frac{2}{b}\Theta(q_{+}-q)\Theta(q-q_{-})
\end{equation}
where $\Theta(x)$ is a Heaviside step function of argument $x$ ($\Theta(x)$ is
$1$ for $x>0$ and $0$ for $x<0$). We then find that
\begin{equation}
\frac{df_{Q_{b}}(q)}{db}=-\frac{f_{Q_{b}}(q)}{b}+\frac{1}{2b}\left[
\delta(q-q_{+})+\delta(q-q_{-})\right] \label{df/db}
\end{equation}
where $\delta(x)$ denotes a Dirac delta function of argument $x$.

Let us now introduce a general increasing transformation, namely a real
function of $q$, written $G(q)$, which is differentiable and strictly
increasing. The extreme values that $G(Q_{b})$ can take are
\begin{equation}
G_{\pm}\equiv G(q_{\pm}(b)).
\end{equation}
We shall now use $\mathbb{E}[...]$ to denote an expected value over $U$. Then
the variance of $G(Q_{b})$ is $\operatorname*{Var}(G(Q_{b}))=\mathbb{E}\left[
G^{2}(Q_{b})\right]  -\mathbb{E}^{2}\left[  G(Q_{b})\right]  $ and we shall
write this in the shorter notation
\begin{align}
V_{b}  &  =\mathbb{E}\left[  G^{2}(Q_{b})\right]  -\mathbb{E}^{2}\left[
G(Q_{b})\right] \nonumber\\
& \nonumber\\
&  =\overline{G^{2}}-\bar{G}^{2}
\end{align}
noting that $\bar{G}=\mathbb{E}\left[  G(Q_{b})\right]  $ and $\overline{G^{2}
}=\mathbb{E}\left[  G^{2}(Q_{b})\right]  $ are both functions of $b$.

Multiplying Eq. (\ref{df/db}) by $G(Q_{b})$ and integrating gives
\begin{equation}
\frac{d\bar{G}}{db}=-\frac{\bar{G}}{b}+\frac{1}{2b}\left(  G_{+}+G_{-}\right)
.
\end{equation}
Multiplying Eq. (\ref{df/db}) by $G^{2}(Q_{b})$ and integrating gives
\begin{equation}
\frac{d\overline{G^{2}}}{db}=-\frac{\overline{G^{2}}}{b}+\frac{1}{2b}\left(
G_{+}^{2}+G_{-}^{2}\right)  .
\end{equation}
The derivative $\frac{dV_{b}}{db}=\frac{d\left(  \overline{G^{2}}-\bar{G}
^{2}\right)  }{db}$ can be written as
\begin{align}
\frac{dV_{b}}{db}  & =\frac{\left[  \left(  \frac{G_{+}^{2}+G_{-}^{2}}%
{2}\right)  -\left(  \frac{G_{+}+G_{-}}{2}\right)  ^{2}\right]  -\left(
\overline{G^{2}}-\bar{G}^{2}\right)  }{b}\nonumber\\
& \nonumber\\
& \quad+\frac{\left(  \bar{G}-\frac{G_{+}+G_{-}}{2}\right)  ^{2}}{b}.
\end{align}
We note that:
\begin{enumerate}
\item $\left(  \frac{G_{+}^{2}+G_{-}^{2}}{2}\right)  -\left(  \frac
{G_{+}+G_{-}}{2}\right)  ^{2}$ is the variance of a distribution containing
two Dirac delta functions that are located at $G_{\pm}$ that each have weight
$1/2$. This is the maximum possible variance of a random variable that takes
values in the interval $[G_{-},G_{+}]$.

\item $\overline{G^{2}}-\bar{G}^{2}$ is the variance of the random variable
$G(Q_{b})$ whose distribution is non-zero in the entire interval $[G_{-}
,G_{+}]$.

\item $\left(  \bar{G}-\frac{G_{+}+G_{-}}{2}\right)  ^{2}$ is non-negative.
\end{enumerate}

From (1) and (2) we have $\left(  \frac{G_{+}^{2}+G_{-}^{2}}{2}\right)
-\left(  \frac{G_{+}+G_{-}}{2}\right)  ^{2}>\overline{G^{2}}-\bar{G}^{2}$ and
because of this and (3) we have that $\frac{dV_{b}}{db}>0$. Integrating this
inequality from $b=1$ to $b=2$ yields $V_{2}>V_{1}$ or explicitly
\begin{equation}
\operatorname*{Var}(G(Q_{2}))>\operatorname*{Var}(G(Q_{1})).\label{Var ineq}
\end{equation}
This result tells us that for \textit{any} increasing transformation, which we
write as $G(q)$, the variance of the transformed version of $Q_{2}$, namely
$G(Q_{2})$, will always exceed the variance of the transformed version of
$Q_{1}$, namely $G(Q_{1})$. In other words, no increasing transformation can
homogenize the variances of the two forms of the conditional distribution
$f_{Q|E}(q|e)$.

\subsection{Figure 3e and 3f}

We now consider the situation illustrated in Figs. 3e and 3f and described in
the associated text.

The distribution of $Q^{\ast}$, conditional on the value of $E^{\ast}$,
generally written as $f_{Q^{\ast}|E^{\ast}}(q^{\ast}|e^{\ast})$, takes two
different forms according to whether $e^{\ast}<3/5$ or $e^{\ast}\geq3/5$. We
write these two different forms in this part of the appendix as $f_{Q^{\ast
}|E^{\ast} }(q^{\ast}|e^{\ast}<3/5)$ and $f_{Q^{\ast}|E^{\ast}}(q^{\ast
}|e^{\ast} \geq3/5)$, respectively. In Fig. 3e we plot $f_{Q^{\ast}|E^{\ast}
}(q^{\ast}|e^{\ast}<3/5)$ as a function of $q^{\ast}$, as given by
\[
f_{Q^{\ast}|E^{\ast}}(q^{\ast}|e^{\ast}<3/5)=\left\{
\begin{array}
[c]{lll}
0, &  & \text{for }0\leq q^{\ast}<\frac{1}{10}\\
&  & \\
\frac{5}{4}, &  & \text{for }\frac{1}{10}\leq q^{\ast}<\frac{9}{10}\\
&  & \\
0, &  & \text{for }\frac{9}{10}\leq q^{\ast}\leq1
\end{array}
\right.
\]
while in Fig. 3f we plot
\[
f_{Q^{\ast}|E^{\ast}}(q^{\ast}|e^{\ast}\geq3/5)=\left\{
\begin{array}
[c]{lll}
\frac{5}{2}, &  & \text{for }0\leq q^{\ast}<\frac{1}{10}\\
&  & \\
\frac{5}{8}, &  & \text{for }\frac{1}{10}\leq q^{\ast}<\frac{9}{10}\\
&  & \\
\frac{5}{2}, &  & \text{for }\frac{9}{10}\leq q^{\ast}\leq1.
\end{array}
\right.
\]

We have 
\begin{align*}
h(Q^{\ast}|e^{\ast})  & =-\int_{0}^{1}f_{Q^{\ast}|E^{\ast}}(q^{\ast}|e^{\ast
})\\
& \\
& \quad\times\log_{2}\left[  f_{Q^{\ast}|E^{\ast}}(q^{\ast}|e^{\ast})\right]
dq^{\ast}%
\end{align*}
and hence
\begin{equation}
h(Q^{\ast}|e^{\ast}<3/5)=-\int_{1/10}^{9/10}\frac{5}{4}\log_{2}\left(
\frac{5}{4}\right)  dq^{\ast}=2-\log_{2}(5)
\end{equation}
while
\begin{equation}%
\begin{array}
[c]{rcl}%
h(Q^{\ast}|e^{\ast}\geq3/5) & = & -\int_{0}^{1/10}\frac{5}{2}\log_{2}\left(
\frac{5}{2}\right)  dq^{\ast}\\
&  & \\
&  & -\int_{1/10}^{9/10}\frac{5}{8}\log_{2}\left(  \frac{5}{8}\right)
dq^{\ast}\\
&  & \\
&  & -\int_{9/10}^{1}\frac{5}{2}\log_{2}\left(  \frac{5}{2}\right)  dq^{\ast
}\\
&  & \\
& = & 2-\log_{2}(5).
\end{array}
\end{equation}
Thus, the two different forms of $f_{Q^{\ast}|E^{\ast}}(q^{\ast}|e^{\ast})$
lead to the same value of the entropy of $Q^{\ast}$, when $E^{\ast}$ is
conditioned to lie in two different ranges. This example is an illustration of
$h(Q^{\ast}|e^{\ast})$ being independent of the value of $e^{\ast}$.

Let us now consider $h(E^{\ast}|q^{\ast})$. Because of Eq. (\ref{same}) we
have $h(E^{\ast}|q^{\ast})=-\int_{0}^{1}f_{Q^{\ast}|E^{\ast}}(q^{\ast}
|e^{\ast})\log_{2}\left[  f_{Q^{\ast}|E^{\ast}}(q^{\ast}|e^{\ast})\right]
de^{\ast}$ hence
\begin{equation}%
\begin{array}
[c]{rcl}%
h(E^{\ast}|q^{\ast}) & = & -\int_{0}^{3/5}f_{Q^{\ast}|E^{\ast}}(q^{\ast
}|e^{\ast}<3/5)\\
&  & \\
&  & \times\log_{2}\left[  f_{Q^{\ast}|E^{\ast}}(q^{\ast}|e^{\ast
}<3/5)\right]  de^{\ast}\\
&  & \\
&  & -\int_{3/5}^{1}f_{Q^{\ast}|E^{\ast}}(q^{\ast}|e^{\ast}\geq3/5)\\
&  & \\
&  & \times\log_{2}\left[  f_{Q^{\ast}|E^{\ast}}(q^{\ast}|e^{\ast}%
\geq3/5)\right]  de^{\ast}.
\end{array}
\end{equation}
This leads to
\begin{align}
&  h(E^{\ast}|q^{\ast})\nonumber\\
& \nonumber\\
&  =\left\{
\begin{array}
[c]{ll}%
-\int_{3/5}^{1}\frac{5}{2}\log_{2}\left(  \frac{5}{2}\right)  de^{\ast}, &
\text{for }0\leq q^{\ast}<\frac{1}{10}\\
& \\
\left.
\begin{array}
[c]{l}%
-\int_{0}^{3/5}\frac{5}{4}\log_{2}\left(  \frac{5}{4}\right)  de^{\ast}\\
\\
-\int_{3/5}^{1}\frac{5}{8}\log_{2}\left(  \frac{5}{8}\right)  de^{\ast}%
\end{array}
\right\}  , & \text{for }\frac{1}{10}\leq q^{\ast}<\frac{9}{10}\\
& \\
-\int_{3/5}^{1}\frac{5}{2}\log_{2}\left(  \frac{5}{2}\right)  de^{\ast}, &
\text{for }\frac{9}{10}\leq q^{\ast}\leq1
\end{array}
\right.
\end{align}
i.e.,
\begin{equation}
h(E^{\ast}|q^{\ast})=\left\{
\begin{array}
[c]{ll}%
1-\log_{2}\left(  5\right)  , & \text{for }0\leq q^{\ast}<\frac{1}{10}\\
& \\
\frac{9}{4}-\log_{2}(5), & \text{for }\frac{1}{10}\leq q^{\ast}<\frac{9}{10}\\
& \\
1-\log_{2}\left(  5\right)  , & \text{for }\frac{9}{10}\leq q^{\ast}\leq1.
\end{array}
\right.  \label{h(E*;q*)}%
\end{equation}
We explicitly see that the entropy of $E^{\ast}$, when $Q^{\ast}$ is
conditioned to take different particular values, exhibits variation.

The \textit{conditional entropy} of $E^{\ast}$ given $Q^{\ast}$ is defined as
\begin{equation}
h(E^{\ast}|Q^{\ast})=\int_{0}^{1}h(E^{\ast}|q^{\ast})\tilde{f}_{Q^{\ast}
}(q^{\ast})dq^{\ast}.
\end{equation}
Since $Q^{\ast}$ is uniformly distributed over $0$ to $1$, it follows that
$\int_{0}^{1}h(E^{\ast}|q^{\ast})\tilde{f}_{Q^{\ast}} (q^{\ast})dq^{\ast}$ is
given by
\begin{align*}
h(E^{\ast}|Q^{\ast})  &  =\int_{0}^{1}h(E^{\ast}|q^{\ast})dq^{\ast}\\
& \\
&  =\frac{2}{10}\left[  1-\log_{2}\left(  5\right)  \right]  +\frac{8}
{10}\left[  \frac{9}{4}-\log_{2}(5)\right] \\
& \\
&  =2-\log_{2}(5).
\end{align*}
This value of $h(E^{\ast}|Q^{\ast})$ coincides with the value of $h(Q^{\ast
}|E^{\ast})$ ($\equiv h(Q^{\ast}|e^{\ast})$). Thus, despite the variation in
the entropy of $E^{\ast}$ that is exhibited when $Q^{\ast}$ is conditioned to
take different particular values, we nevertheless have $h(E^{\ast}|Q^{\ast
})=h(Q^{\ast}|E^{\ast})$, as required by the basic results concerning mutual information.

A last point we shall make concerns what happens when $E^{\ast}$ takes its
mutual information-maximising form. In this case the distribution of $E^{\ast
}$ is uniform over $0$ to $1$. We can calculate the information gained about
$E^{\ast}$ when $Q^{\ast}$ is observed to have the particular value $q^{\ast}
$, that we write as $I(E^{\ast};q^{\ast})$. We find from Eq. (\ref{h(E*;q*)})
that
\begin{align}
I(E^{\ast};q^{\ast})  &  =h(E^{\ast})-h(E^{\ast}|q^{\ast})\nonumber\\
& \nonumber\\
&  =\left\{
\begin{array}
[c]{lll}
\log_{2}\left(  5\right)  -1, &  & \text{for }0\leq q^{\ast}<\frac{1}{10}\\
&  & \\
\log_{2}(5)-\frac{9}{4}, &  & \text{for }\frac{1}{10}\leq q^{\ast}<\frac{9
}{10}\\
&  & \\
\log_{2}\left(  5\right)  -1, &  & \text{for }\frac{9}{10}\leq q^{\ast}\leq1.
\end{array}
\right. \nonumber\\
&
\end{align}
This result explicitly shows that $I(E^{\ast};q^{\ast})$ depends on $q^{\ast}
$. In other words, we gain more information about the value of $E^{\ast}$ when
we observe some values of $Q^{\ast}$ compared with other values. \newpage

\bibliographystyle{ieeetr}
\bibliography{BIB_PeckWaxman}

\providecommand{\noopsort}[1]{}\providecommand{\singleletter}[1]{#1}%
\begin{thebibliography}{10}

\bibitem{PeckWaxman2018}
J.~R. Peck and D.~Waxman, ``What is adaptation and how should it be
  measured?,'' {\em J. Theor. Biol.}, vol.~447, pp.~190--198, 2018.

\bibitem{Shannon}
C.~E. Shannon, ``A mathematical theory of communication,'' {\em Bell Syst.
  Tech.J.}, vol.~27, pp.~623--656, 1948.

\bibitem{CoverThomas}
T.~M. Cover and J.~A. Thomas, {\em Elements of Information Theory}.
\newblock Wiley-Blackwell, New York, 1991.

\bibitem{MacKay}
D.~J.~C. MacKay, {\em Information Theory, Inference and Learning Algorithms.
  Sixth Printing}.
\newblock Cambridge University Press, New York, 2007.

\bibitem{Brillinger}
D.~R. Brillinger, ``Some data analyses using mutual information,'' {\em Braz.
  J. Probab. Stat.}, vol.~18, pp.~163--182, 2004.

\bibitem{Adesso}
G.~Adesso, N.~Datta, M.~Hall, and T.~Sagawa, ``Shannon’s information theory
  70 years on: applications in classical and quantum physics,'' {\em J. Phys.
  Math. Theor.}, vol.~52, p.~320201, 2019.

\bibitem{Kraskov}
A.~Kraskov, H.~St{\"o}gbauer, and P.~Grassberger, ``Estimating mutual
  information,'' {\em Phys. Rev. E}, vol.~69, p.~066138, 2004.

\bibitem{haigh2013probability}
J.~Haigh, {\em Probability Models}.
\newblock Springer, London, 2013.

\bibitem{Casella}
G.~Casella and R.~L. Berger, {\em Statistical Inference}.
\newblock Thomson Learning, Pacific Grove CA, 2002.

\bibitem{Vasicek}
O.~Vasicek, ``A test for normality based on sample entropy,'' {\em J. R. Stat.
  Soc. Ser. B Methodol.}, vol.~38, pp.~54--59, 1976.

\bibitem{Chen}
B.~Chen, J.~Wang, H.~Zhao, and J.~C. Principe, ``Insights into entropy as a
  measure of multivariate variability,'' {\em Entropy}, vol.~18, p.~196, 2016.

\bibitem{Baran}
T.~Baran, F.~Barbaros, A.~G{\"u}l, and G.~O. G{\"u}l, ``Entropy as a variation
  of information for testing the goodness of fit,'' {\em Water Resour. Manag.},
  vol.~32, pp.~5151--5168, 2018.

\bibitem{Laughlin}
S.~Laughlin, ``A simple coding procedure enhances a neuron's information
  capacity,'' {\em Z. für Naturforschung C}, vol.~36, pp.~910--912, 1981.

\bibitem{Tkacik1}
G.~Tkacik, C.~G. Callan, and W.~Bialek, ``Information capacity of genetic
  regulatory elements,'' {\em Phys. Rev. E}, vol.~78, p.~011910, 2008.

\bibitem{Tkacik2}
G.~Tkacik, C.~G. Callan, and W.~Bialek, ``Information flow and optimization in
  transcriptional regulation,'' {\em Proc. Natl. Acad. Sci.}, vol.~105,
  pp.~12265--12270, 2008.

\bibitem{Gonzalez}
R.~C. Gonzalez and R.~E. Woods, {\em Digital Image Processing, 4th Edition}.
\newblock Pearson, New York, 2018.

\bibitem{Szostak}
J.~W. Szostak, ``Functional information: Molecular messages,'' {\em Nature},
  vol.~423, pp.~689--689, 2003.

\bibitem{Hazen}
R.~M. Hazen, P.~L. Griffin, J.~M. Carothers, and J.~W. Szostak, ``Functional
  information and the emergence of biocomplexity,'' {\em Proc. Natl. Acad.
  Sci.}, vol.~104, pp.~8574--8581, 2007.

\bibitem{Adami2000}
C.~Adami, C.~Ofria, and T.~C. Collier, ``Evolution of biological complexity,''
  {\em Proc. Natl. Acad. Sci.}, vol.~97, pp.~4463--4468, 2000.

\bibitem{Adami2002}
C.~Adami, ``What is complexity?,'' {\em BioEssays}, vol.~24, pp.~1085--1094,
  2002.

\bibitem{Brown}
J.~H. Brown and B.~A. Maurer, ``Body size, ecological dominance and cope’s
  rule,'' {\em Nature}, vol.~324, pp.~248--250, 1986.

\bibitem{Alroy}
J.~Alroy, ``Cope’s rule and the dynamics of body mass evolution in north
  american fossil mammals,'' {\em Science}, vol.~280, pp.~731--734, 1998.

\bibitem{Morand}
S.~Morand and R.~Poulin, ``Density, body mass and parasite species richness of
  terrestrial mammals,'' {\em Evol. Ecol.}, vol.~12, pp.~717--727, 1998.

\bibitem{van_Kampen}
N.~G. van Kampen, {\em Stochastic Processes in Physics and Chemistry, 3rd
  edition}.
\newblock North Holland, Amsterdam, 2007.

\bibitem{Barton}
G.~Barton, {\em Elements of Green's Functions and Propagation: Potentials,
  Diffusion, and Waves}.
\newblock Oxford University Press, Oxford, 1989.

\end{thebibliography}

\end{document}